\documentclass[%
 reprint,
 amsmath,amssymb,
 aps,
prb,
]{revtex4-1}

\usepackage{graphicx}% Include figure files
\usepackage{dcolumn}% Align table columns on decimal point
\usepackage{bm}% bold math
\newcommand{\R}{\mathbb{R}}
\newcommand{\C}{\mathbb{C}}

\newcommand{\ie}{\textit{i.e.}\/, }
\newcommand{\eg}{\textit{e.g.}\/, }
\newcommand{\cf}{\textit{cf.}\/, }

\providecommand*{\mrm}[1]{\mathrm{#1}}
\providecommand*{\unit}[1]{\ensuremath{\mrm{\,#1}}}
\providecommand*{\eu}{\ensuremath{\mrm{e}}}
\providecommand*{\iu}{\ensuremath{\mrm{i}}}

\renewcommand{\Re}{\ensuremath{\mrm{Re}}}	% The LaTeX standard is not ISO!
\renewcommand{\Im}{\ensuremath{\mrm{Im}}}	% The LaTeX standard is not ISO!

\newcommand{\maximize}{\mrm{maximize}}

\newcommand{\subto}{\mrm{subject\ to}}

\def\XXint#1#2#3{{\setbox0=\hbox{$#1{#2#3}{\int}$}
     \vcenter{\hbox{$#2#3$}}\kern-.5\wd0}}

\newtheorem{theorem}{\em Theorem}[section]

\begin{document}

%\preprint{APS/123-QED}

%\title{Optimal plasmonic resonances of the sphere in lossy media}% Force line breaks with \\
%\title{On the optimal plasmonic resonances in lossy media: \\ Multipole resonances of the sphere}
\title{Optimal plasmonic multipole resonances of a sphere in lossy media}

\author{Sven~Nordebo}
 \email{sven.nordebo@lnu.se}
\affiliation{%
 Department of Physics and Electrical Engineering, \\ Linn\ae us University, 351 95 V\"{a}xj\"{o}, Sweden.
}%

\author{Gerhard~Kristensson}
 \email{gerhard.kristensson@eit.lth.se}
\affiliation{%
Department of Electrical and Information Technology, Lund University, Box 118, 221 00 Lund, Sweden.
}%

\author{Mohammad~Mirmoosa}
\email{mohammad.mirmoosa@aalto.fi}
\author{Sergei~Tretyakov}
\email{sergei.tretyakov@aalto.fi}
\affiliation{
 Department of Electronics and Nanoengineering, \\ Aalto University, P.O. Box 15500, FI-00076 Aalto, Finland.
}%

\date{\today}% It is always \today, today,

\begin{abstract}
%Fundamental upper bounds of power absorbed and scattered by optimally tuned multipole-resonant plasmonic spheres embedded in a lossy surrounding medium.
Fundamental upper bounds are given for the %optimal 
plasmonic multipole absorption and scattering of a rotationally invariant  
dielectric sphere embedded in a lossy surrounding medium.  A specialized Mie theory is developed for this purpose and when combined with 
the corresponding generalized optical theorem, an optimization problem is obtained which is explicitly 
solved by straightforward analysis. In particular, the absorption cross section is a concave quadratic form in the related Mie (scattering) parameters and 
the convex scattering cross section can be maximized by using a Lagrange multiplier constraining the absorption to be non-negative.
%An asymptotic analysis is then carried out to characterize the corresponding resonances of small homogeneous spheres.
For the homogeneous sphere, the Weierstrass preparation theorem is used to establish the existence and the uniqueness of the plasmonic singularities 
and explicit asymptotic expressions are given for the dipole and the quadrupole. 
It is shown that the optimal passive material for multipole absorption and scattering of a small homogeneous dielectric sphere embedded in a dispersive medium 
is given approximately as the complex conjugate and the real part of the corresponding pole positions, respectively.
Numerical examples are given to illustrate the theory, including a comparison with the plasmonic dipole and quadrupole resonances obtained 
in gold, silver and aluminum nanospheres based on some specific Brendel-Bormann (BB) dielectric models for these metals. % of gold, silver and aluminum. 
Based on these BB-models, it is interesting to note that the metal spheres can be tuned to optimal absorption at a particular size at a particular frequency.
\end{abstract}

\maketitle

%\tableofcontents

\section{Introduction}

The absorption and scattering of light by small particles have many interesting applications in plasmonics concerning \eg 
plasmon waveguides, aperture arrays, extraordinary transmission, perfect lenses, artificial magnetism, 
and surface-enhanced biological sensing with molecular monolayer spectroscopy, only to mention a few \cite{Maier2007}.
There are also new emerging applications and ideas emanating from the (in principle) unlimited-power reflection, absorption, and emission 
that is associated with high modal orders and super resolution effects \cite{Valagiannopoulos+etal2015,Maslovski+etal2016,Valagiannopoulos+Tretyakov2016}.
In the analysis of these problems the background medium is usually assumed to be lossless. 
This assumption is usually made for simplicity, but it also gives very strong results such as the optical theorem for a small particle
of arbitrary shape, and which is given solely in terms of the wavelength of the incident light and the polarizability of the particle \cite[pp.~71 and 140]{Bohren+Huffman1983}.

In some of these applications, however, it may also sometimes be necessary, or even critically important, to consider the losses in the surrounding medium.
The plasmonic resonances in small particles, and in particular the resonances associated with high modal orders, will be severely limited by the presence of internal, as well as external losses. 
Many of the substances used in optics such as the polymeric media are usually considered to be transparent to visible light \cite{Progelhof+etal1971},
but there are also studies concerning doped PMMA showing significant losses \cite{AlTaay+etal2015}. Another interesting application area 
is with light in biological tissue \cite{Duck1990} and the use of gold nanoparticles for plasmonic photothermal therapy \cite{Huang+etal2008}.
Hence, there is a motivation to develop new theory that can be used to evaluate the impact of external losses in these applications.

Interestingly as it turns out, it is a non-trivial task to develop a general theory on the scattering and absorption for small particles  
in a lossy surrounding medium and the existing results are typically given only for spheres \cite{Mundy+etal1974,Chylek1977,Bohren+Gilra1979,Lebedev+etal1999,Sudiarta+Chylek2001,Durant+etal2007a}. 
An important example is Bohren's optical theorem for a spherical particle in an absorbing medium where the extinction cross section
is defined in a way to be consistent with the power loss that can be physically observed at a detector when the particle is placed 
between the source and the detector \cite[Eq.~(11) on p.~218]{Bohren+Gilra1979}. 
However, this definition of extinction cross section diverts from the more common definition made in \cite[Eq.~(3.19) on p.~70]{Bohren+Huffman1983}
or \cite[Eq. (7) on p.~1276]{Sudiarta+Chylek2001} but which is not necessarily non-negative when the surrounding medium is lossy. 
Nevertheless, as we will see in this paper, the latter definition is very useful as a mathematical device to derive the optimal absorption of a sphere inside a lossy medium.

As has been indicated above, there are major difficulties associated with the definition of cross sections when there are losses in the surrounding medium (even for the sphere).
Consider \eg the simple fact that the power intensity of the plane wave impinging on the particle will depend on the spatial variable.
Hence, the fundamental difficulties in general are that the absorption in the surrounding medium will depend on the geometry of the scatterer 
and the cross sections will depend on the chosen reference point (or origin) for the plane wave.

The purpose of this paper is to develop a specialized theory with rigorous bounds on the multipole scattering and absorption for 
a rotationally invariant sphere embedded in an infinite isotropic lossy medium. 
For simplicity, explicit formulas are given only for the electrical multipoles of non-magnetic, 
dielectric or metallic spheres which are common in plasmonic applications, but the magnetic cases can be treated similarly.

The rest of the paper is organized as follows: 
In Section \ref{eq:Electrodynamics} we derive the basic formulas regarding the electrodynamics of a rotationally invariant sphere embedded in a lossy medium.
The general theory is then exploited in Section \ref{sect:optplasmres} for derivation of fundamental upper bounds on scattering and absorption, as well as for characterization of 
optimal plasmonic resonances of the homogeneous sphere.
The theory is then illustrated in Section \ref{sect:numexamples} with numerical examples concerning the optimal absorption of gold, silver and aluminium nanospheres. 
The paper is finally concluded with a summary.

\section{Electrodynamics of the sphere}\label{eq:Electrodynamics}

\subsection{Notation and conventions}
%The notations and conventions used in this paper are as follows.
The classical Maxwell's equations are considered with electric and magnetic field intensities $\bm{E}$ and $\bm{H}$ given in 
SI-units \cite{Jackson1999,Kristensson2016,Osipov+Tretyakov2017}. The time convention for time harmonic fields (phasors) is given by $\eu^{-\iu\omega t}$
where $\omega$ is the angular frequency and $t$ the time. 
Let $\mu_0$, $\epsilon_0$, $\eta_0$ and $\mrm{c}_0$ denote the permeability, the permittivity, the wave impedance and
the speed of light in vacuum, respectively, where $\eta_0=\sqrt{\mu_0/\epsilon_0}$ and $\mrm{c}_0=1/\sqrt{\mu_0\epsilon_0}$.
The wavenumber of vacuum is given by $k_0=\omega\sqrt{\mu_0\epsilon_0}$.
The wavenumber of a homogeneous and isotropic medium with relative permeability $\mu$ and permittivity $\epsilon$ is given by $k=k_0\sqrt{\mu\epsilon}$
and the wavelength $\lambda$ is defined by $k\lambda=2\pi$.
The wave impedance of the same medium is given by $\eta_0\eta$ where $\eta=\sqrt{\mu/\epsilon}$ is the relative wave impedance.
In the following, we will consider only non-magnetic, homogeneous and isotropic materials, and hence $\mu=1$ from now on. 
The spherical coordinates are denoted by $(r,\theta,\phi)$, the corresponding unit vectors $(\hat{\bm{r}},\hat{\bm{\theta}},\hat{\bm{\phi}})$,
and the radius vector $\bm{r}=r\hat{\bm{r}}$.
The regular spherical Bessel functions, the Neumann functions, the spherical Hankel functions of the first kind 
and the corresponding Riccati-Bessel functions \cite{Kristensson2016} are denoted $\mrm{j}_l(z)$, $\mrm{y}_l(z)$, $\mrm{h}_l^{(1)}(z)=\mrm{j}_l(z)+\iu\mrm{y}_l(z)$,
$\psi_l(z)=z\mrm{j}_l(z)$ and $\xi_l(z)=z\mrm{h}_l^{(1)}(z)$, respectively, all of order $l$.
Finally, the real and the imaginary parts, and the complex conjugate of a complex number $z$ are denoted 
$\Re\left\{z\right\}$, $\Im\left\{z\right\}$ and $z^*$, respectively.

\subsection{General Mie theory for a lossy background}
The Mie theory gives the solution to Maxwell's equations for a plane wave impinging on a homogeneous sphere in terms of the multipole expansion
of spherical vector waves, see \eg \cite{Bohren+Huffman1983,Kristensson2016}.
The definition of the spherical vector waves and a description of some of their most important properties used in this paper
are given in Appendix \ref{sect:spherical}.

We consider the scattering of the electromagnetic field due to a 
homogeneous and isotropic dielectric sphere of radius $a$, relative permittivity $\epsilon$, 
and wavenumber $k=k_0\sqrt{\epsilon}$.
The background medium is assumed to be homogeneous and isotropic and is
characterized by the relative permittivity $\epsilon_\mrm{b}$ and the associated wavenumber $k_\mrm{b}=k_0\sqrt{\epsilon_\mrm{b}}$.
Throughout the analysis in this paper, both materials are assumed to be passive, either lossy or lossless, and hence
$\Im\{\epsilon\}\geq 0$ as well as $\Im\{\epsilon_\mrm{b}\}\geq 0$. % unless explicitly stated otherwise.
It is also assumed that neither $\epsilon$ nor $\epsilon_\mrm{b}$ can reside at the branch-cut of the square root which is defined as the negative part of the real axis.
%The branch-cut of the square root is taken as usual as the negative part of the real line, and it is assumed that neither $\epsilon$ nor $\epsilon_\mrm{b}$
%can be real valued and negative.

The incident and the scattered fields for $r>a$ are expressed as in \eqref{eq:EHsphdef} with multipole coefficients $a_{\tau ml}$ and $f_{\tau ml}$
for regular and outgoing spherical vector waves, respectively, 
and the interior field for $r<a$ is similarly expressed using regular spherical vector waves with multipole coefficients $a_{\tau ml}^\mrm{int}$.
By matching the tangential fields at the boundary of radius $a$, it can be shown that
\begin{eqnarray}\label{eq:btaumldef}
f_{\tau ml}=t_{\tau l}a_{\tau ml}, \\
a_{\tau ml}^\mrm{int}=r_{\tau l}a_{\tau ml},\label{eq:ataumldef}
\end{eqnarray}
for $\tau=1,2$, $l=1,2,\ldots$, and $m=-l,\ldots,l$, and where $t_{\tau l}$ and $r_{\tau l}$ are transition matrices for scattering and absorption, 
respectively, see \eg \cite[Eqs.~(4.52) and (4.53) on p.~100]{Bohren+Huffman1983}.
In particular, the electric ($\tau=2$) multipole coefficients are given by
\begin{equation} 
t_{2l} =
\displaystyle  \frac{ m \psi_l(ka)\psi_l^\prime(k_\mrm{b}a)-\psi_l(k_\mrm{b}a)\psi_l^\prime(ka)}
{ \xi_l(k_\mrm{b}a)\psi_l^\prime(ka)-m\psi_l(ka) \xi_l^\prime(k_\mrm{b}a)},  
%\displaystyle  \frac{ \mrm{j}_l(ka) \psi_l^\prime(k_\mrm{b}a) \epsilon - \mrm{j}_l(k_\mrm{b}a) \psi_l^\prime (ka) \epsilon_\mrm{b}}
%{ \mrm{h}_l^{(1)}(k_\mrm{b}a) \psi_l^\prime(ka)\epsilon_\mrm{b}- \mrm{j}_l(ka) \xi_l^\prime(k_\mrm{b}a)\epsilon },  
\label{eq:t2l} 
\end{equation}
 and 
\begin{equation} 
r_{2l} = 
\displaystyle \frac{ -\iu m }
   { \xi_l(k_\mrm{b}a)\psi_l^\prime(ka)-m\psi_l(ka) \xi_l^\prime(k_\mrm{b}a)},   
%\displaystyle  \frac{1}{k_0a}\frac{ \iu\sqrt{\epsilon} }
 %  { -\mrm{h}_l^{(1)}(k_\mrm{b}a) \psi_l^\prime(ka)\epsilon_\mrm{b} + \mrm{j}_l(ka) \xi_l^\prime(k_\mrm{b}a)\epsilon},   
   \label{eq:r2l}
\end{equation}
where $m=\sqrt{\epsilon}/\sqrt{\epsilon_\mrm{b}}$, \cf also \cite[Eqs.~(8.7) and (8.10)  on pp.~420 and 426]{Kristensson2016}.
%and where the Wronskian for the Riccati-Bessel functions have been used to simplify $r_{2l}$
%\cf \cite[Eq.~(8.10) on p.~426]{Kristensson2016}.

Let $\bm{E}_\mrm{i}(\bm r)=\bm{E}_0\eu^{i k_\mrm{b}\hat{\bm{k}}\cdot\bm{r}}$ describe a plane wave
with vector amplitude $\bm{E}_0$ and propagation direction $\hat{\bm{k}}$.
It can be shown that the corresponding multipole expansion coefficients are given by
\begin{equation}
a_{\tau ml}=4\pi\iu^{l-\tau+1}\bm{E}_0\cdot{\bf A}_{\tau ml}^*(\hat{\bm{k}}),
\end{equation}
for $\tau=1,2$, $l=1,2,\ldots$, and $m=-l,\ldots,l$, and where the vector spherical harmonics ${\bf A}_{\tau ml}(\hat{\bm{k}})$
are defined as in Appendix \ref{sect:sphericaldef}, see also \cite[Eq.~(7.28) on p.~375]{Kristensson2016}. 
Based on the sum identities for the vector spherical harmonics derived in Appendix \ref{sect:sumidspheharm}
it is readily seen that
\begin{equation}\label{eq:sumoverm}
\sum_{m=-l}^l \left| a_{\tau ml} \right|^2=2\pi (2l+1)\left| \bm{E}_0\right|^2,
\end{equation}
and where \eqref{eq:sumA2} and \eqref{eq:sumA1} have been used for $\tau=2$ and $\tau=1$, respectively, as well as the relation $\hat{\bm{k}}\cdot\bm{E}_0=0$.
It is noticed that the relation \eqref{eq:sumoverm} is independent of the direction $\hat{\bm{k}}$ of the incoming plane wave.

Although the main focus of this paper is with the homogeneous and isotropic dielectric sphere, 
many of the relevant results developed here are also valid for a rotationally invariant sphere.
With a rotationally invariant sphere the scattering behavior can be described as in \eqref{eq:btaumldef}, \ie with a diagonal transition matrix 
$t_{\tau l}$ which is independent of the azimuthal index $m$. A typical situation is with a layered sphere \cite[Chapter 8.3]{Kristensson2016}
or with a radially inhomogeneous sphere having an index of refraction which depends only on the radial coordinate $r$.%, see \eg \cite{Sihvola}.

\subsection{Rotationally invariant sphere in a lossy medium}
Consider the scattering of a rotationally invariant sphere of volume $\mrm{V}_a$ bounded by the spherical surface $\mrm{S}_a$ of radius $a$ 
and which is embedded inside a lossy (or lossless) infinite, 
homogeneous, isotropic and non-magnetic background medium with complex-valued relative permittivity $\epsilon_\mrm{b}$.
Let $\bm{E}$ and $\bm{H}$ denote the total fields everywhere in $\R^3$, and $\bm{E}_\mrm{i}$ and 
$\bm{E}_\mrm{s}$ the incident and the scattered fields, respectively, so that $\bm{E}=\bm{E}_\mrm{i}+\bm{E}_\mrm{s}$ in the exterior region $\R^3\setminus \overline{\mrm{V}_a}$, and
similarly for the magnetic field. Based on the expansion in spherical vector waves \eqref{eq:EHsphdef} together with \eqref{eq:btaumldef} as well as \eqref{eq:sumoverm}
and the orthogonality relationships  \eqref{eq:orthcrosssph1} and \eqref{eq:orthcrosssph2},
the power absorbed in the scatterer can be expressed in terms of the exterior fields as
\begin{multline}\label{eq:PabsdefII}
P_\mrm{abs}=-\int_{\mrm{S}_a}\frac{1}{2}\Re\left\{\left(\bm{E}_\mrm{i}+\bm{E}_\mrm{s}\right)\times\left(\bm{H}_\mrm{i}+\bm{H}_\mrm{s}\right)^*\right\}\cdot\mrm{d}\bm{S} \\
=\frac{\pi\left| \bm{E}_0\right|^2}{\eta_0}\Im\left\{ \left(\sqrt{\epsilon_\mrm{b}}\right)^*\sum_{\tau=1}^2\sum_{l=1}^\infty  (2l+1) \right. \\
\left[\int_{\mrm{S}_a}{\bf v}_{\tau ml}(k_\mrm{b}\bm{r})\times{\bf v}_{\bar{\tau} ml}^*(k_\mrm{b}\bm{r})\cdot\mrm{d}\bm{S} \right. \\
+t_{\tau l}\int_{\mrm{S}_a}{\bf u}_{\tau ml}(k_\mrm{b}\bm{r})\times{\bf v}_{\bar{\tau} ml}^*(k_\mrm{b}\bm{r})\cdot\mrm{d}\bm{S} \\
+ t_{\tau l}^*\int_{\mrm{S}_a}{\bf v}_{\tau ml}(k_\mrm{b}\bm{r})\times{\bf u}_{\bar{\tau} ml}^*(k_\mrm{b}\bm{r})\cdot\mrm{d}\bm{S} \\
\left.\left. +\left|t_{\tau l}\right|^2\int_{\mrm{S}_a}{\bf u}_{\tau ml}(k_\mrm{b}\bm{r})\times{\bf u}_{\bar{\tau} ml}^*(k_\mrm{b}\bm{r})\cdot\mrm{d}\bm{S}
\right]
\right\},
\end{multline}
where ${\bf v}_{\tau ml}(k_\mrm{b}\bm{r})$ and ${\bf u}_{\tau ml}(k_\mrm{b}\bm{r})$ are the regular and the outgoing spherical vector waves, respectively, and
$\bar{\tau}$ denotes the dual index, etc., \cf Appendix \ref{sect:sphericaldef}.
In the order of the terms appearing in \eqref{eq:PabsdefII}, \ie the constant, the linear and the quadratic forms in $t_{\tau l}$, the absorbed power can be 
interpreted as $P_\mrm{abs}=P_\mrm{i}+P_\mrm{ext}-P_\mrm{s}$ where $P_i$ relates to the power lost in the background medium,  $P_\mrm{ext}$ is the extinct power and
$P_\mrm{s}$ the scattered power, respectively \cite[p.~70]{Bohren+Huffman1983}. 
Strictly speaking,  $P_\mrm{ext}$ can be interpreted as the extinct power $P_\mrm{ext}=P_\mrm{abs}+P_\mrm{s}$
only when the surrounding medium is lossless and $P_\mrm{i}=0$, \cf \cite[Eq.~(3.20) on p.~70]{Bohren+Huffman1983}. 
In the general lossy case, $P_\mrm{ext}$ is interpreted here merely as a mathematical device
to calculate the absorption, see also \eg \cite{Bohren+Gilra1979,Lebedev+etal1999,Sudiarta+Chylek2001,Durant+etal2007a} for a comprehensive discussion on this topic.

The corresponding cross sections are obtained by the normalization $C=P/I_\mrm{i}$ where 
$I_\mrm{i}=\left| \bm{E}_0 \right|^2\Re\{\sqrt{\epsilon_\mrm{b}}\}/2\eta_0$ is the intensity of the plane wave at the origin $\bm{r}=\bm{0}$.
Based on the orthogonality relationships  \eqref{eq:orthcrosssph1} and \eqref{eq:orthcrosssph2},
the absorption cross section for a particular electric ($\tau=2$) multipole index $l$ can now be expressed in terms of the following generalized optical theorem
\begin{multline}\label{eq:Cabs2ldef}
C_{\mrm{abs},2l}
=-C_{\mrm{s},2l}+C_{\mrm{ext},2l}+C_{\mrm{i},2l}  \\
=\frac{2\pi(2l+1)}{\left| k_\mrm{b}\right|^2}\left(a_{2l}\left|t_{2 l} \right|^2+2\Re\left\{b_{2l}t_{2 l}\right\}+c_{2l}
\right),
\end{multline}
where
\begin{equation}\label{eq:a2ldef}
a_{2l}=-\Im\left\{\frac{k_\mrm{b}^*}{\Re\{k_\mrm{b}\}} \xi_l^\prime(k_\mrm{b}a)\xi_l^*(k_\mrm{b}a)\right\},
\end{equation}
\begin{multline}\label{eq:b2ldef}
b_{2l}=\frac{1}{2\iu}\left(-\frac{k_\mrm{b}^*}{\Re\{k_\mrm{b}\}} \xi_l^\prime(k_\mrm{b}a)\psi_l^*(k_\mrm{b}a) \right. \\
\left. +\frac{k_\mrm{b}}{\Re\{k_\mrm{b}\}} {\psi_l^\prime}^*(k_\mrm{b}a)\xi_l(k_\mrm{b}a)\right),
\end{multline}
and
\begin{equation}\label{eq:c2ldef}
c_{2l}=-\Im\left\{\frac{k_\mrm{b}^*}{\Re\{k_\mrm{b}\}} \psi_l^\prime(k_\mrm{b}a)\psi_l^*(k_\mrm{b}a)\right\},
\end{equation}
see also \cite[Eq. (7) on p.~1276]{Sudiarta+Chylek2001}. Observe that the notation used in \eqref{eq:a2ldef} through \eqref{eq:c2ldef} should not be confused with \eqref{eq:btaumldef} and \eqref{eq:ataumldef}.
%Magnetic dipoles ($\tau=1$) can be treated similarly.
It is noted that the coefficients $a_{2l}$ and $c_{2l}$ are real-valued whereas the coefficients $b_{2l}$ are complex-valued.
Furthermore, it is seen that the scattering cross section is given by 
\begin{equation}\label{eq:Cs2ldef}
C_{\mrm{s},2l}=-\frac{2\pi(2l+1)}{\left| k_\mrm{b}\right|^2}a_{2l}\left|t_{2 l} \right|^2,
\end{equation}
and since Poynting's theorem asserts that the scattered power must be non-negative in a passive medium, it follows immediately from \eqref{eq:Cs2ldef} that $a_{2l}<0$.
By employing the Wronskians of the spherical Bessel functions ($\mrm{j}_l\mrm{y}_l^\prime-\mrm{j}_l^\prime\mrm{y}_l=1/z^2$) 
and of the Riccati-Bessel functions ($\psi_l\xi_l^\prime-\psi_l^\prime\xi_l=\iu$), it can be shown that for a lossless medium with $\Im\{k_\mrm{b}\}=0$,
the coefficients defined in \eqref{eq:a2ldef} through \eqref{eq:c2ldef} become $a_{2l}=-1$,  $b_{2l}=-1/2$ and $c_{2l}=0$, yielding
\begin{equation}
C_{\mrm{abs},2l}=\frac{2\pi(2l+1)}{k_\mrm{b}^2}\left(-\left|t_{2 l} \right|^2-\Re\{t_{2 l}\}\right),
\end{equation}
which is in agreement with the classical optical theorem, see \eg \cite[p.~421]{Kristensson2016}
or \cite[Eqs.~(7.295) and (7.297) on pp.~465-466]{Osipov+Tretyakov2017}.

\subsection{Homogeneous sphere in a lossy medium}
The absorption of a homogeneous sphere can also be calculated directly from the internal fields via
Poynting's theorem 
\begin{equation}
P_\mrm{abs}=\frac{1}{2}\omega\epsilon_0\Im\{\epsilon\}\int_{\mrm{V}_{a}} \left| \bm{E} \right|^2\mrm{d}v,
\end{equation}
where $\mrm{V}_{a}$ denotes the spherical volume of radius $a$.
Due to the orthogonality of the spherical vector waves over a spherical volume as expressed in \eqref{eq:vorthogonal2} through \eqref{eq:W2ldef}, the absorption cross section can 
be evaluated as
\begin{multline}\label{eq:Cabssphexpdef}
C_\mrm{abs}=\frac{P_\mrm{abs}}{I_\mrm{i}}=\frac{k_0\Im\{\epsilon\}}{\left| E_0\right|^2\Re\{\sqrt{\epsilon_\mrm{b}}\}}\int_{\mrm{V}_a} \left| \bm{E} \right|^2\mrm{d}v  \\
 =\frac{2\pi k_0\Im\{\epsilon\}}{\Re\{\sqrt{\epsilon_\mrm{b}}\}}\sum_{l=1}^{\infty}\sum_{\tau=1}^2  (2l+1) W_{\tau l}(k,a)\left| r_{\tau l} \right|^2,
\end{multline}
where $W_{\tau l}(k,a)=\int_{\mrm{V}_a}\left|{\bf v}_{\tau ml}(k\bm{r})\right|^2\mrm{d} v$ are the volume integrals of the regular spherical vector waves.
Note that in \eqref{eq:Cabssphexpdef}, the relations \eqref{eq:ataumldef} and \eqref{eq:sumoverm} have also been employed.
For the electric multipoles ($\tau=2$),  $W_{2 l}(k,a)$ is given by \eqref{eq:W1ldef} and \eqref{eq:W2ldef} and can be expressed explicitly as
\begin{multline}\label{eq:W2lexpr}
W_{2l}(k,a)=\frac{a^2}{2l+1}\frac{1}{\Im\!\left\{k^2\right\}} 
 \Im\!\left\{k\left[ (l+1)\mrm{j}_{l}(ka) \mrm{j}_{l-1}^*(ka) \right.\right.  \\
 \left.\left. +l\mrm{j}_{l+2}(ka) \mrm{j}_{l+1}^*(ka) \right]\right\},
\end{multline}
and which are based on the adequate Lommel integrals for spherical Bessel functions with complex-valued arguments,
\cf also \cite[Eq.~(A.15) on p.~11]{Nordebo+etal2017a}.

\section{Optimal plasmonic resonances of the sphere}\label{sect:optplasmres}
\subsection{Optimal absorption of the rotationally invariant sphere}
Consider the problem of maximizing the absorption cross section for a single electric ($\tau=2$) multipole index $l$ based on \eqref{eq:Cabs2ldef}. 
Since $a_{2l}<0$,  the absorption cross section $C_{\mrm{abs},2l}$ is a strictly concave quadratic form in $t_{2 l}$, hence possessing
a unique maximum. By differentiating \eqref{eq:Cabs2ldef} with respect to the complex conjugate of the variable $z=t_{2l}$ and solving for stationarity 
\begin{equation}
\frac{\partial}{\partial z^*}\left(a_{2l}zz^*+b_{2l}z+b_{2l}^*z^*+c_{2l}\right)=a_{2l}z+b_{2l}^*=0,
\end{equation}
the following optimal Mie coefficient is obtained
\begin{equation}
t_{2l}=-\frac{b_{2l}^*}{a_{2l}},
\end{equation}
yielding the optimal absorption cross section for the rotationally invariant sphere
\begin{equation}\label{eq:Cabsopt}
C_{\mrm{abs},2l}^\mrm{opt}=\frac{\pi(2l+1)}{2\left|k_\mrm{b}\right|^2}\left(-\frac{4\left| b_{2l} \right|^2}{a_{2l}}+4c_{2l}\right).
\end{equation}
%Note that the same result is obtained when differentiating with respect to $z$ and where the differential operators are defined as in \cite[p.~12]{Greene+Krantz2002}.

For a lossless medium with $\Im\{k_\mrm{b}\}=0$, we have $a_{2l}=-1$,  $b_{2l}=-1/2$ and $c_{2l}=0$, giving the optimal solution
$t_{2 l}=-1/2$, and 
\begin{equation}\label{eq:CabsoptLL}
C_{\mrm{abs},2 l}^\mrm{opt}=\frac{\pi(2l+1)}{2k_\mrm{b}^2},
\end{equation}
in agreement with the classical theory for a lossless medium, see \eg \cite[Eq.~(16) on p.~937]{Tretyakov2014} for the case $l=1$.

\subsection{Optimal scattering of the rotationally invariant sphere}
Consider the problem of maximizing the scattering cross section for a single electric ($\tau=2$) multipole index $l$ based on \eqref{eq:Cs2ldef}.
Since $a_{2l}<0$, the scattering cross section $C_{\mrm{s},2l}$ is a strictly convex quadratic form in $t_{2 l}$, and the
additional convex constraint $C_{\mrm{abs},2l}\geq 0$ based on  \eqref{eq:Cabs2ldef} is needed to get a valid solution.
With $z=t_{2l}$, the problem can be formulated equivalently as
\begin{equation}\label{eq:Csoptprobdef}
\begin{array}{llll}
	& \maximize & & \displaystyle zz^* \vspace{0.2cm} \\    
	& \subto & &  a_{2l}zz^*+b_{2l}z+b_{2l}^*z^*+c_{2l} =0,
\end{array}
\end{equation}
and where the inequality constraint has been replaced by an equality constraint ($C_{\mrm{abs},2l}=0$) since the maximum of a convex function over a convex set will always occur
at a boundary point (active constraint). The Lagrange function for \eqref{eq:Csoptprobdef} is given by
\begin{equation}\label{eq:Lagrangefun}
L(z)=zz^*+\alpha\left( a_{2l}zz^*+b_{2l}z+b_{2l}^*z^*+c_{2l} \right),
\end{equation}
where $\alpha$ is the real-valued multiplier. The Lagrange condition for an optimal solution is 
\begin{equation}\label{eq:Lagrangefunsol}
\frac{\partial L}{\partial z^*}=z+\alpha\left( a_{2l}z+b_{2l}^*  \right) =0,
\end{equation}
and the solution can be written as
\begin{equation}\label{eq:zsolLagrange}
z=\frac{-\alpha b_{2l}^*}{1+\alpha a_{2l}}=\beta b_{2l}^*,
\end{equation}
where $\beta=-\alpha/(1+\alpha a_{2l})$ is a real-valued parameter.
Insertion of \eqref{eq:zsolLagrange} into the quadratic constraint in \eqref{eq:Csoptprobdef} gives the condition for stationarity
\begin{equation}\label{eq:stationarity}
\beta^2a_{2l}\left|b_{2l}\right|^2+2\beta\left|b_{2l}\right|^2+c_{2l}=0.
\end{equation}
The  maximizing solution is readily found as $t_{2 l}=\beta b_{2l}^*$ where
\begin{equation}\label{eq:ttaulopts}
\beta=-\frac{1}{a_{2l}}+\sqrt{\frac{1}{a_{2l}^2}-\frac{c_{2l}}{a_{2l}\left|b_{2l}\right|^2}},
\end{equation}
and the optimal scattering cross section for the rotationally invariant sphere is
\begin{equation}\label{eq:Csopt}
C_{\mrm{s},2 l}^\mrm{opt}=\frac{2\pi(2l+1)}{\left|k_\mrm{b}\right|^2}(-a_{2l})\left|b_{2l}\right|^2\beta^2.
\end{equation}

For a lossless medium with  $\Im\{k_\mrm{b}\}=0$, we have $a_{2l}=-1$,  $b_{2l}=-1/2$ and $c_{2l}=0$ giving the optimal solution
$t_{2 l}=-1$, and 
\begin{equation}\label{eq:CsoptLL}
C_{\mrm{s},2 l}^\mrm{opt}=\frac{2\pi(2l+1)}{k_\mrm{b}^2},
\end{equation}
in agreement with the classical theory for a lossless medium, see \eg \cite[Eq.~(17) on p.~938]{Tretyakov2014} for the case $l=1$.

\subsection{Asymptotic analysis for small homogeneous spheres}
An asymptotic analysis of \eqref{eq:Cabssphexpdef} and \eqref{eq:Cs2ldef} is carried out to find approximate expressions for the corresponding 
optimal permittivity $\epsilon_\mrm{opt}$ of the homogeneous sphere when the electrical size $k_0a$ is small.
For this purpose, the following power series expansions of the spherical Bessel functions  is employed
\begin{equation}\label{eq:besseljlexpr}
\mrm{j}_l(z)=\sum_{k=0}^\infty A_{kl}z^{l+2k},
\end{equation}
and for the spherical Neumann functions
\begin{equation}\label{eq:besselylexpr}
\mrm{y}_l(z)=\sum_{k=0}^{l} B_{kl}z^{-l-1+2k}+{\cal O}\{z^{l+1}\},
\end{equation}
and the spherical Hankel functions of the first kind
\begin{equation}\label{eq:besselhlexpr}
\mrm{h}_l^{(1)}(z)=\iu\sum_{k=0}^{l} B_{kl}z^{-l-1+2k}+A_{0l}z^l+{\cal O}\{z^{l+1}\},
\end{equation}
where $A_{kl}=(-\frac{1}{2})^k/k!(2l+2k+1)!!$ and $B_{kl}=-(\frac{1}{2})^k(2l-2k-1)!!/k!$ \cf \cite[Eqs.~(10.53.1) and (10.53.2), respectively]{Olver+etal2010}
and where ${\cal O}\{\cdot\}$ denotes the big ordo defined in \cite[p.~4]{Olver1997}. 
The power series expansions of $\psi_l(z)$, $\xi_l(z)$, $\psi_l^\prime(z)$ and $\xi_l^\prime(z)$ are readily obtained from \eqref{eq:besseljlexpr} and \eqref{eq:besselhlexpr}.
%Note that $\psi_l^\prime(z)$ and $\xi_l^\prime(z)$ share the same regularity properties as of $\mrm{j}_l(z)$ and $\mrm{h}_l^{(1)}(z)$ and that their
%power series expansions are readily obtained from \eqref{eq:besseljlexpr} and \eqref{eq:besselhlexpr}, respectively.

Consider the Mie series coefficient  $r_{2l}$ given by \eqref{eq:r2l} and extend the fraction using the factor $\iu\epsilon_\mrm{b}(k_\mrm{b}a)^l(ka)^{-l}$ to get
\begin{equation} 
r_{2l}  %= \displaystyle  \frac{1}{k_0a}\frac{\sqrt{\epsilon} (k_\mrm{b}a)^{l+1}(ka)^{-l}}{f_l(\epsilon,\sqrt{\epsilon_\mrm{b}},k_0a)} 
=\frac{\left(\sqrt{\epsilon_\mrm{b}}\right)^{l+1}}{\left(\sqrt{\epsilon}\right)^{l-1}}\frac{1}{f_l(\epsilon,\sqrt{\epsilon_\mrm{b}},k_0a)}, \label{eq:r2lexpanded}
\end{equation}
where
\begin{multline} \label{eq:Fldef}
f_l(\epsilon,\sqrt{\epsilon_\mrm{b}},k_0a)= \iu(k_\mrm{b}a)^{l}\xi_l(k_\mrm{b}a) (ka)^{-l}\psi_l^\prime(ka)\epsilon_\mrm{b} \\
-\iu (ka)^{-l-1}\psi_l(ka) (k_\mrm{b}a)^{l+1}\xi_l^\prime(k_\mrm{b}a)\epsilon \\
=\frac{l}{2l+1}\epsilon+\frac{l+1}{2l+1}\epsilon_\mrm{b}+{\cal O}\{(k_0a)^2\},
\end{multline}
and where the order relation is found by considering the corresponding power series expansions.
Note that the combinations $z^{l}\xi_l(z)$, $z^{-l}\psi_l^{\prime}(z)$, $z^{-l-1}\psi_l(z)$ and $z^{l+1}\xi_l^\prime(z)$ are whole analytic functions.
Hence, it can be concluded that the function $f_l(\epsilon,\sqrt{\epsilon_\mrm{b}},k_0a)$ defined in \eqref{eq:Fldef} is an analytic function in all of its three 
arguments $(\epsilon,\sqrt{\epsilon_\mrm{b}},k_0a)$.
The function $f_l$ is analytic in $\epsilon$ since $z^{-l}\psi_l^\prime(z)$ and $z^{-l-1}\psi_l(z)$ are even functions.

An asymptotic analysis of \eqref{eq:r2lexpanded}  shows that % based on the power series expansions above
\begin{multline}\label{eq:r2lasymptoticexpansion}
r_{2l}=\frac{2l+1}{l}\frac{\left(\sqrt{\epsilon_\mrm{b}}\right)^{l+1}}{\left(\sqrt{\epsilon}\right)^{l-1}}
\frac{1}{\epsilon+\frac{l+1}{l}\epsilon_\mrm{b}+(k_0a)^2C_l(k_0a,\epsilon,\epsilon_\mrm{b})} \\
\frac{}{+\iu(k_0a)^{2l+1}D_l +{\cal O}\{(k_0a)^{2l+2}\}},
\end{multline}
where $C_l(k_0a,\epsilon,\epsilon_\mrm{b})$ is a polynomial function of $(k_0a,\epsilon,\epsilon_\mrm{b})$ 
with terms having even order in $k_0a$ ranging from 0 up to $2l-2$ and real-valued coefficients, and
\begin{equation}\label{eq:Dldef}
D_l=\frac{l+1}{l}\frac{1}{(2l+1)!!}\frac{1}{(2l-1)!!}\left(\sqrt{\epsilon_\mrm{b}}\right)^{2l+1}(\epsilon_\mrm{b}-\epsilon).
\end{equation}
A detailed study shows that for the electric dipole we have
\begin{multline}\label{eq:r21asymptoticexpansion}
r_{21}=3\frac{\epsilon_\mrm{b}}
{\epsilon+2\epsilon_\mrm{b}+(k_0a)^2 \left(\epsilon_\mrm{b}^2-\frac{9}{10}\epsilon_\mrm{b}\epsilon -\frac{1}{10}\epsilon^2 \right)} \\
\frac{}{+\iu(k_0a)^{3}\frac{2}{3}\epsilon_\mrm{b}\sqrt{\epsilon_\mrm{b}}(\epsilon_\mrm{b}-\epsilon)+{\cal O}\{(k_0a)^{4}\}},
\end{multline}
and for the quadrupole
\begin{multline}\label{eq:r22asymptoticexpansion}
r_{22}=\frac{5}{2\sqrt{\epsilon}}\frac{\epsilon_\mrm{b}\sqrt{\epsilon_\mrm{b}}}
{\epsilon+\frac{3}{2}\epsilon_\mrm{b}+(k_0a)^2 \left(\frac{1}{4}\epsilon_\mrm{b}^2-\frac{5}{28}\epsilon_\mrm{b}\epsilon -\frac{1}{14}\epsilon^2 \right)} \\
\frac{}{+(k_0a)^4 \left(\frac{1}{16}\epsilon_\mrm{b}^3-\frac{1}{14}\epsilon_\mrm{b}^2\epsilon +\frac{1}{144}\epsilon_\mrm{b}\epsilon^2 + \frac{1}{504}\epsilon^3 \right)} \\
\frac{}{+\iu(k_0a)^{5}\frac{1}{30}\epsilon_\mrm{b}^2\sqrt{\epsilon_\mrm{b}}(\epsilon_\mrm{b}-\epsilon)+{\cal O}\{(k_0a)^{6}\}}.
\end{multline}
%Note that both $\epsilon$ as well as $\epsilon_\mrm{b}$ are in general complex-valued throughout this analysis.

 \subsection{The plasmonic singularities of the homogeneous sphere}\label{sect:plasmonsingularity}
%From the power series expansions leading to \eqref{eq:r2lasymptoticexpansion}
%it is found that the Mie coefficient $r_{\tau l}$ can be written
When the background permittivity $\epsilon_\mrm{b}$ is fixed, it follows from \eqref{eq:r2lexpanded} and \eqref{eq:Fldef} 
that the Mie coefficient $r_{2 l}$ can be written
\begin{equation}\label{eq:r2lWeier}
r_{2l}=\frac{2l+1}{l}\frac{\left(\sqrt{\epsilon_\mrm{b}}\right)^{l+1}}{\left(\sqrt{\epsilon}\right)^{l-1}}\frac{1}{f_l(w,k_0a)}
\end{equation}
where $f_l(w,z)$ is an analytic function in the complex variables $w$ and $z$, of the form
\begin{equation}\label{eq:fWeierdef}
f_l(w,z)=w+{\cal O}\{z^2\},
\end{equation}
and where $w=\epsilon+\frac{l+1}{l}\epsilon_\mrm{b}$ and  $z=k_0a$.

The following theorem by Weierstrass \cite[Theorem 7.5.1]{Hormander1983} can now be used to establish the existence and the uniqueness of a single pole of $r_{2 l}$ with the 
property $\epsilon_{\mrm{p},l}(k_0a)\rightarrow -\frac{l+1}{l}\epsilon_\mrm{b}$ as $k_0a\rightarrow 0$.
We will refer to $\epsilon_{\mrm{p},l}(k_0a)$ as {\em the plasmonic multipole singularity of the sphere}.
\begin{theorem}\label{th:weierstrass} {\rm (The Weierstrass preparation theorem)}\\
Let $f(w,z)$ be an analytic function of $(w,z)\in\C\times\C$ in a neighborhood of $(0,0)$ such that
\begin{equation}
\left\{\begin{array}{l}
\displaystyle f=\frac{\partial f}{\partial w}=\ldots=\frac{\partial^{n-1} f}{\partial w^{n-1}}=0, \vspace{0.2cm} \\
\displaystyle \frac{\partial^{n} f}{\partial w^{n}}\neq 0,
\end{array}\right.
\end{equation} 
at $(0,0)$. Then there is a unique factorization
\begin{equation}
f(w,z)=a(w,z)\left(w^n+b_{n-1}(z)w^{n-1}+\ldots + b_0(z) \right),
\end{equation}
where $b_j(z)$ and $a(w,z)$ are analytic in a neighborhood of $0$ and $(0,0)$, respectively,
$a(0,0)\neq 0$ and $b_j(0)=0$.
\end{theorem}

Here, it is seen from \eqref{eq:fWeierdef} that
\begin{equation}\label{eq:weirstrassprereq}
\left\{\begin{array}{l}
f_l=0,  \vspace{0.2cm} \\
\displaystyle \frac{\partial f_l}{\partial w}=1,
\end{array}\right.
\end{equation}
at $(w,z)=(0,0)$. It follows then from Theorem \ref{th:weierstrass} that there is a unique factorization
\begin{equation}\label{eq:ffact}
f_l(w,z)=a(w,z)\left(w+b_0(z) \right),
\end{equation}
where $a(w,z)$ and $b_0(z)$ are analytic in a neighborhood of $(0,0)$ and $0$, respectively,
and $a(0,0)\neq 0$ and $b_0(0)=0$. 
From \eqref{eq:fWeierdef} follows also that $f_l(w,0)=w$ and hence that $a(w,0)=1$.
The factorization \eqref{eq:ffact} can now be written
\begin{equation}\label{eq:ffact2}
f_l(w,k_0a)=a(w,k_0a)\left(\epsilon- \epsilon_{\mrm{p},l}(k_0a) \right),
\end{equation}
where
\begin{equation}\label{eq:epsilonpk0adef}
\epsilon_{\mrm{p},l}(k_0a)=-\frac{l+1}{l}\epsilon_\mrm{b}-b_0(k_0a),
\end{equation}
and where $b_0(k_0a)\rightarrow 0$ as  $k_0a\rightarrow 0$. This establishes the existence and the uniqeness of the pole $\epsilon_{\mrm{p},l}(k_0a)$ as stated above.
Since $a(w,z)$ is a continuous function with $a(w,0)=1$, it is furthermore seen that
\begin{equation}
f_l(w,k_0a)\sim \epsilon- \epsilon_{\mrm{p},l}(k_0a) 
\end{equation}
as $k_0a\rightarrow 0$ and where the symbol $\sim$ indicates an asymptotic approximation in the sense of \cite[p.~4]{Olver1997}.
Finally, the asymptotics of the Mie coefficient $r_{2 l}$ can be written
\begin{equation}\label{eq:r2lWeierasymptot}
r_{2l}\sim \frac{2l+1}{l}\frac{\left(\sqrt{\epsilon_\mrm{b}}\right)^{l+1}}{\left(\sqrt{\epsilon}\right)^{l-1}}\frac{1}{\epsilon- \epsilon_{\mrm{p},l}(k_0a)}
\end{equation}
as $k_0a\rightarrow 0$. It is also observed that $\epsilon_{\mrm{p},l}(k_0a)$ resides in the lower complex half plane $\Im\{\epsilon\}<0$ when the surrounding media is passive, 
\cf \eqref{eq:epsilonpk0adef} with $\Im\{\epsilon_\mrm{b}\}> 0$.

An asymptotic analysis of the equation $f_l(w,k_0a)=0$ to leading orders in $k_0a$ reveals the pole structure of the electric multipole.
In general, the pole structure based on \eqref{eq:r2lasymptoticexpansion} and \eqref{eq:Dldef} is given by
\begin{multline}\label{eq:multipolestructure}
\epsilon_{\mrm{p},l}(k_0a)=-\frac{l+1}{l}\epsilon_\mrm{b}+(k_0a)^{2}F_l(k_0a,\epsilon,\epsilon_\mrm{b}) \\
-\iu(k_0a)^{2l+1}\epsilon_\mrm{b}^{l+1}\sqrt{\epsilon_\mrm{b}}\frac{l+1}{l^2}\frac{1}{((2l-1)!!)^2}+{\cal O}\{(k_0a)^{2l+2}\},
\end{multline}
where $F_l(k_0a,\epsilon_\mrm{b})$ is a polynomial function of $(k_0a,\epsilon_\mrm{b})$ 
with terms having even order in $k_0a$ ranging from 0 up to $2l-2$ and real-valued coefficients.
A detailed study based on \eqref{eq:r21asymptoticexpansion} and \eqref{eq:r22asymptoticexpansion} gives the following pole expression for the dipole
\begin{multline}\label{eq:poleasexprdipole}
\epsilon_{\mrm{p},1}(k_0a)=-2\epsilon_\mrm{b}-\frac{12}{5}\epsilon_\mrm{b}^2(k_0a)^2 \\ -\iu 2\epsilon_\mrm{b}^2\sqrt{\epsilon_\mrm{b}}(k_0a)^3+{\cal O}\{(k_0a)^4\},
\end{multline}
and for the quadrupole
\begin{multline}\label{eq:poleasexprquadrupole}
\epsilon_{\mrm{p},2}(k_0a)=-\frac{3}{2}\epsilon_\mrm{b}-\frac{5}{14}\epsilon_\mrm{b}^2(k_0a)^2 - \frac{65}{392}\epsilon_\mrm{b}^3(k_0a)^4 \\
-\iu \frac{1}{12}\epsilon_\mrm{b}^3\sqrt{\epsilon_\mrm{b}}(k_0a)^5+{\cal O}\{(k_0a)^6\}.
\end{multline}

\subsection{Optimal permittivity of the homogeneous sphere}\label{sect:optdielhomsphere}

To maximize the absorption cross section \eqref{eq:Cabssphexpdef} with respect to the permittivity $\epsilon$ we consider
the normalized absorption cross section $Q_{\mrm{abs},2 l}$ for a particular electric multipole
\begin{multline}\label{eq:Qabssphexpdef}
Q_{\mrm{abs},2 l}=\frac{C_{\mrm{abs},2 l}}{\pi a^2}
 =\frac{2 k_0a\Im\{\epsilon\}}{\Re\{\sqrt{\epsilon_\mrm{b}}\}}  (2l+1) \frac{W_{2 l}(k,a)}{a^3}\left| r_{2 l} \right|^2 \\
 =\frac{2}{\Re\{\sqrt{\epsilon_\mrm{b}}\}} \Im\!\left\{\sqrt{\epsilon}\left[ (l+1)\mrm{j}_{l}(ka) \mrm{j}_{l-1}^*(ka) \right.\right. \\
\left.\left.  +l\mrm{j}_{l+2}(ka) \mrm{j}_{l+1}^*(ka) \right]\right\}\left|r_{2l} \right|^2,
\end{multline}
where  \eqref{eq:W2lexpr} has been used.
An asymptotic analysis of \eqref{eq:Qabssphexpdef} using \eqref{eq:besseljlexpr} and \eqref{eq:r2lWeierasymptot} yields
\begin{multline}\label{eq:Qabsasymptot}
Q_{\mrm{abs},2 l} \\ \sim
 (k_0a)^{2l-1}\frac{2(l+1)(2l+1)}{l^2((2l-1)!!)^2}\frac{\left|\epsilon_\mrm{b}\right|^{l+1}}{\Re\{\sqrt{\epsilon_\mrm{b}}\}}\frac{\Im\{\epsilon\}}{\left|\epsilon- \epsilon_{\mrm{p},l}(k_0a) \right|^2},
\end{multline}
for small $k_0a$, and where it is observed that the singular factor $1/\left(\sqrt{\epsilon}\right)^{l-1}$ in \eqref{eq:r2lWeierasymptot} cancel due to the corresponding regularity of $W_{2 l}(k,a)$.
The function $Q_{\mrm{abs},2 l}$ is of the form $F(\epsilon)=\Im\{\epsilon\}/|\epsilon- \epsilon_{\mrm{p},l}(k_0a)|^2$ which has a local maximum for $\Im\{\epsilon\}>0$ at
\begin{equation}\label{eq:optepsabs2l}
\epsilon_{\mrm{opt},l}(k_0a)=\epsilon_{\mrm{p},l}^*(k_0a),
\end{equation}
\cf \eg \cite[Sect.~2.5, Eqs.~(15) through (17)]{Nordebo+etal2017a}. Hence, the expression \eqref{eq:optepsabs2l} gives an approximation of
the optimal permittivity for multipole absorption of small dielectric spheres embedded in lossy media.
 
Based on \eqref{eq:poleasexprdipole} and \eqref{eq:poleasexprquadrupole} we can now immediately assess that the corresponding optimal permittivity
for the dipole absorption is given by
\begin{multline}\label{eq:poleasexprdipoleopt}
\epsilon_{\mrm{opt},1}(k_0a) =-2\epsilon_\mrm{b}^*-\frac{12}{5}\epsilon_\mrm{b}^{*2}(k_0a)^2 \\ +\iu 2\epsilon_\mrm{b}^{*2}\sqrt{\epsilon_\mrm{b}^*}(k_0a)^3+{\cal O}\{(k_0a)^4\},
\end{multline}
and for the quadrupole absorption
\begin{multline}\label{eq:poleasexprquadrupoleopt}
\epsilon_{\mrm{opt},2}(k_0a)=-\frac{3}{2}\epsilon_\mrm{b}^*-\frac{5}{14}\epsilon_\mrm{b}^{*2}(k_0a)^2 - \frac{65}{392}\epsilon_\mrm{b}^{*3}(k_0a)^4 \\
+\iu \frac{1}{12}\epsilon_\mrm{b}^{*3}\sqrt{\epsilon_\mrm{b}^*}(k_0a)^5+{\cal O}\{(k_0a)^6\}.
\end{multline}

Similarly, to maximize the scattering cross section defined in \eqref{eq:Cs2ldef}, it is observed that
$C_{\mrm{s},2 l}$ is proportional to the squared Mie coefficient $\left| t_{2 l}\right|^2$. An asymptotic analysis of \eqref{eq:t2l} for small $k_0a$ shows that
\begin{multline}\label{eq:t2lsim1}
t_{2l}\sim \iu (k_0a)^{2l+1}\frac{l+1}{l}\frac{1}{(2l+1)!!}\frac{1}{(2l-1)!!}\left(\sqrt{\epsilon_\mrm{b}}\right)^{2l+1} \\
\frac{\epsilon-\epsilon_\mrm{b}}{\epsilon-\epsilon_{\mrm{p},l}(k_0a)},
\end{multline}
where $\epsilon_{\mrm{p},l}(k_0a)$ is the same pole as defined in \eqref{eq:ffact2} and \eqref{eq:epsilonpk0adef} above, and hence that
\begin{equation}\label{eq:t2lsim2}
\left|t_{2l}\right|^2\sim A\frac{\left|\epsilon-\epsilon_\mrm{b}\right|^2}{\left|\epsilon-\epsilon_{\mrm{p},l}(k_0a)\right|^2},
\end{equation}
where $A$ is a constant. To maximize \eqref{eq:t2lsim2} for $\Im\{\epsilon\}\geq 0$ and small $k_0a$ as well as with small losses $\Im\{\epsilon_\mrm{b}\}$, it is readily seen
that the distance function $\left|\epsilon-\epsilon_\mrm{b}\right|$ can be neglected and the optimal plasmonic resonance for multipole scattering by small dielectric spheres in lossy media
is approximately given by
\begin{equation}\label{eq:optepsscatt2l}
\epsilon_{\mrm{opt},l}(k_0a)=\Re\{\epsilon_{\mrm{p},l}(k_0a)\}.
\end{equation}

It is noted that the expressions \eqref{eq:optepsabs2l} through \eqref{eq:poleasexprquadrupoleopt}, and similarly \eqref{eq:optepsscatt2l}
give asymptotic expansions of the permittivity of a small dielectric sphere yielding optimal absorption and scattering, respectively,
and which explicitly takes the background loss into account via the complex-valued parameter $\epsilon_\mrm{b}$. 
This generalizes previous results for a lossless background given \eg by 
%\cite[Eq.~(18) through (24) with a small correction $3/5\rightarrow 12/5$]{Tretyakov2014} 
\cite[Eq.~(18) through (24) on p.~938]{Tretyakov2014} 
and \cite[Eq.~(11) and (14) on p.~3]{Tzarouchis+etal2016}.
It is finally noted that the homogeneous sphere represents a subclass of spherical objects embraced by the rotationally invariant sphere,
and the optimality of the homogeneous sphere must hence be bounded by \eqref{eq:Cabsopt} and \eqref{eq:Csopt}.

%and which is accurate up to the third order in $k_\mrm{b}a$, \cf \cite[Eq.~(11) and (14) on p.~3]{Tzarouchis+etal2016} and \cite{Tretyakov2014}
%(note that the expression \eqref{eq:optdielspheresol} contains a small correction of the corresponding formula given in \cite[Eq.~(22) and (24) on p.~938]{Tretyakov2014}).

%And I have rigorously characterized the optimal (in the sense of asymptotic approximate ÓoptimalÓ absorption or scattering of a small
%sphere as explained in the text) dielectric material of a homogeneous sphere, see
%eg (56) in the attached paper, which is the result for an optimal dipole resonance giving (approximately) optimal dipole absorption
%of a small, homogeneous sphere. This result generalizes the theory given in your equations (18)-(23) in your paper, but here we can see
%exactly how we should deal with a complex valued (lossy) background permittivity \epsilon_b. Note that in your equation (22) the factor 3/5 should 
%in fact be 12/5 as you can also see in Aris paper (Phys Rev B 94 2016) in their eq (11).

\section{Numerical examples}\label{sect:numexamples}

%alpha1=2*2*pi*ftest*(1/c0)*imag(sqrt(epsilon_b1));
%log10(alpha1/100) % absorption coefficient in cm^(-1)
%log10(alpha2/100)
%log10(alpha3/100)
%ans =-3.9422
%ans =2.0578
%ans =4.0573
    
The theory developed in this paper is discussed based on a few numerical examples and illustrated in Figures \ref{fig:matfig101} through \ref{fig:matfig103} below.
In particular, we are relating here to applications in optics where the the absorption coefficient is given by $\alpha=2k_0\Im\{\sqrt{\epsilon_\mrm{b}}\}$
and the skin-depth is $\alpha^{-1}$, \cf  \cite[p.~314-315]{Jackson1999}.
As a reference, in \cite[Table 3.2 on pp.~47--49]{Duck1990} is given a comprehensive summary of published data regarding the 
absorption coefficient of biological tissue at optical frequencies. Even though this data is very diverse, one can argue
that the skin-depth of tissue in the visible light is in the order of $\alpha^{-1}=10^{-1}$\unit{cm}.
Another class of materials which is important in optics are the polymeric media such as PMMA which usually can be considered as transparent
in the optical regime, \cf \cite{Progelhof+etal1971}. However, some papers report large absorption coefficients, and in particular for doped PMMA films
with skin-depths as small as $\alpha^{-1}=10^{-2}$\unit{cm} for visible light \cite{AlTaay+etal2015}.
The skin-depth of pure water is approximately $\alpha^{-1}=10^{4}$\unit{cm} for visible light \cite[p.~315]{Jackson1999}.
The numerical examples given below have been chosen to cover this range of losses, and even though the refractive indices are somewhat different in various
applications, the real part of $\epsilon_\mrm{b}$ is not so critical in these comparisons and we have therefore consistently chosen $\Re\{\epsilon_\mrm{b}\}=1$.

In Figure \ref{fig:matfig101} is shown the optimal normalized absorption cross section $Q_{\mrm{abs},2l}^\mrm{opt}=C_{\mrm{abs},2l}^\mrm{opt}/\pi a^2$ 
where $C_{\mrm{abs},2l}^\mrm{opt}$ is given by \eqref{eq:Cabsopt}, and plotted here as a function of the electrical size $k_0a$ 
for different loss factors $\epsilon_\mrm{b}^{\prime\prime}=\Im\{\epsilon_\mrm{b}\}=10^{-9},10^{-3},10^{-1}$, and multipole orders 
$l=1,2$. At visible light with the wavelength $\lambda=550$\unit{nm}, these loss factors correspond approximately to a
skin-depth $\alpha^{-1}= 10^{4},10^{-2},10^{-4}$ \unit{cm}, respectively.
%and where $\alpha$ is the absorption coefficient given by $\alpha=2k_0\Im\{\sqrt{\epsilon_\mrm{b}}\}$, \cf  \cite[p.~314-315]{Jackson1999}.

\begin{figure}[htb]
\begin{center}
\includegraphics[width=0.48\textwidth]{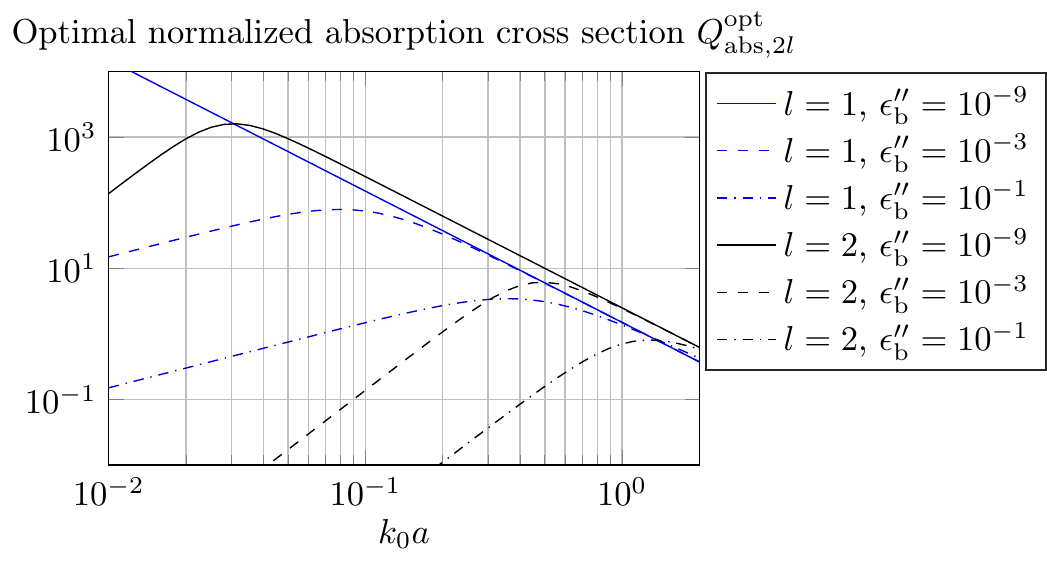}
\end{center}
\vspace{-5mm}
\caption{The optimal normalized absorption cross section $Q_{\mrm{abs},2l}^\mrm{opt}$ for a rotationally invariant sphere in a lossy medium with
$\epsilon_\mrm{b}=1+\iu\epsilon_\mrm{b}^{\prime\prime}$.
%plotted as a function of electrical size $k_0a$ for different loss factors $\epsilon_\mrm{b}^{\prime\prime}$ and multipole orders .
}
\label{fig:matfig101}
\end{figure}

%The maximum of these curves indicate a transition between the {\em lossy quasistatic} and the {\em lossless full dynamic} regimes.
%In other words, the left asymptotes of these curves
%approximate very closely the quasistatic solution for the lossy media where $\epsilon_{\mrm{o},l}(k_0a)\approx-\frac{l+1}{l}\epsilon_\mrm{b}^*$,
%and the right asymptotes approximate very closely the full dynamic solution in the corresponding lossless media where $\epsilon_\mrm{b}^{\prime\prime}\approx 0$.
%In essence, the lossless full dynamic approximation is valid if the electrical size $k_0a$ is sufficiently large at the same time as the external loss 
%$\epsilon_\mrm{b}^{\prime\prime}$ is sufficiently small, and the lossy quasistatic approximation is valid if $k_0a$ is sufficiently small at the same time as 
%$\epsilon_\mrm{b}^{\prime\prime}$ is sufficiently large.
Notice that even though the quadrupole field is potentially more efficient for absorbtion, it is attenuated much more effectively than the dipole field with
increasing external losses or decreasing electrical size.
The optimal scattering cross section \eqref{eq:Csopt} can be investigated similarly and shows a very similar spectrum, only about 4 times larger, 
\cf \eqref{eq:CabsoptLL} and \eqref{eq:CsoptLL}.

\begin{figure}[htb]
\begin{center}
\includegraphics[width=0.48\textwidth]{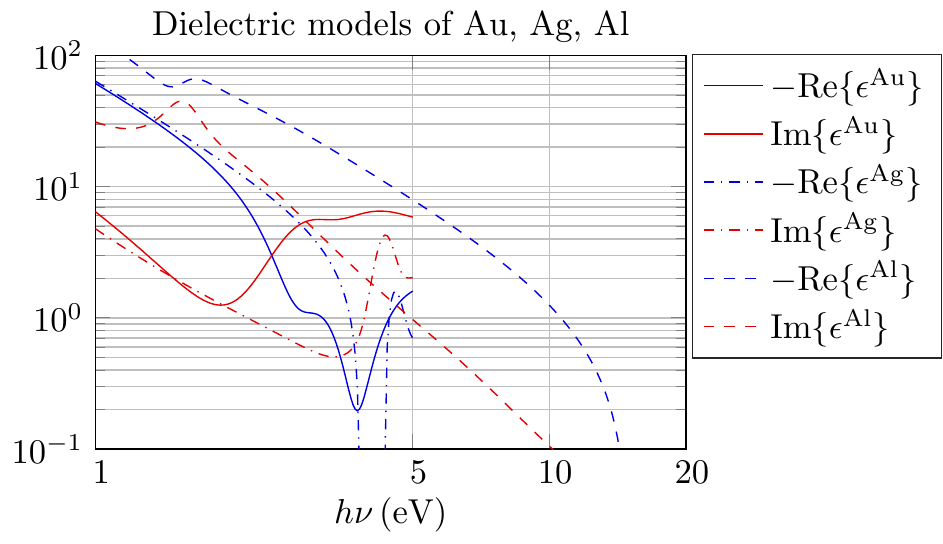}
\end{center}
\vspace{-5mm}
\caption{Permittivities of gold (Au), silver (Ag) and aluminum (Al) according to the Brendel-Bormann model fitted to experimental data \cite{Rakic+etal1998}.}
\label{fig:matfig21}
\end{figure}

\begin{figure}[htb]
\begin{center}
\includegraphics[width=0.48\textwidth]{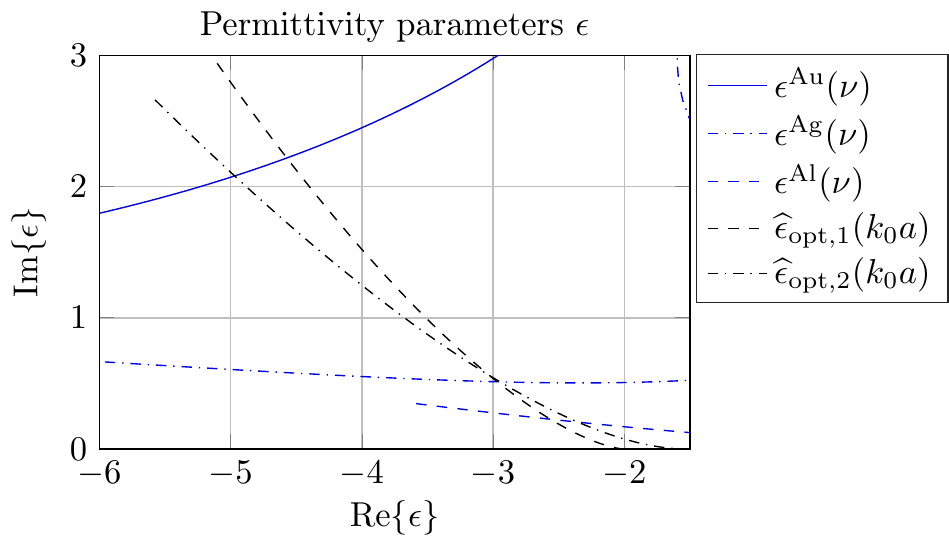}
\end{center}
\vspace{-5mm}
\caption{The dielectric functions $\epsilon^\mrm{Ax}(\nu)$ for gold (Au), silver (Ag) and aluminum (Al) according to the Brendel-Bormann model \cite{Rakic+etal1998}, 
and the asymptotic near-optimal dielectric functions $\widehat{\epsilon}_{\mrm{opt},l}(k_0a)$, $l=1,2$, 
parameterized by frequency $\nu$ and electrical size $k_0a$, respectively, plotted in the complex plane. 
Their intersections give an approximation to an optimal dipole or quadrupole resonance for each metal.}
\label{fig:matfig45}
\end{figure}

%gold
%opt dipole 88.6 nm, k0a=1.0361
%opt quadrupole 165.1 nm, k0a=1.9036
%
%silver 
%opt dipole 38.7 nm, k0a=0.6325
%opt quadrupole 88.0 nm, k0a=1.4398
%
%aluminum
%opt dipole 12.1 nm, k0a=0.4855
%opt quadrupole 28.8 nm, k0a=1.1956

% Dipole peaks at
%Etest=[2.43 3.18 7.85];k0test=Etest*eV/(hbar*c0);lambdatest=10^(9)*2*pi./k0test
%lambdatest =510.2219  389.8866  157.9413

\begin{figure}[htb]
\begin{center}
\includegraphics[width=0.48\textwidth]{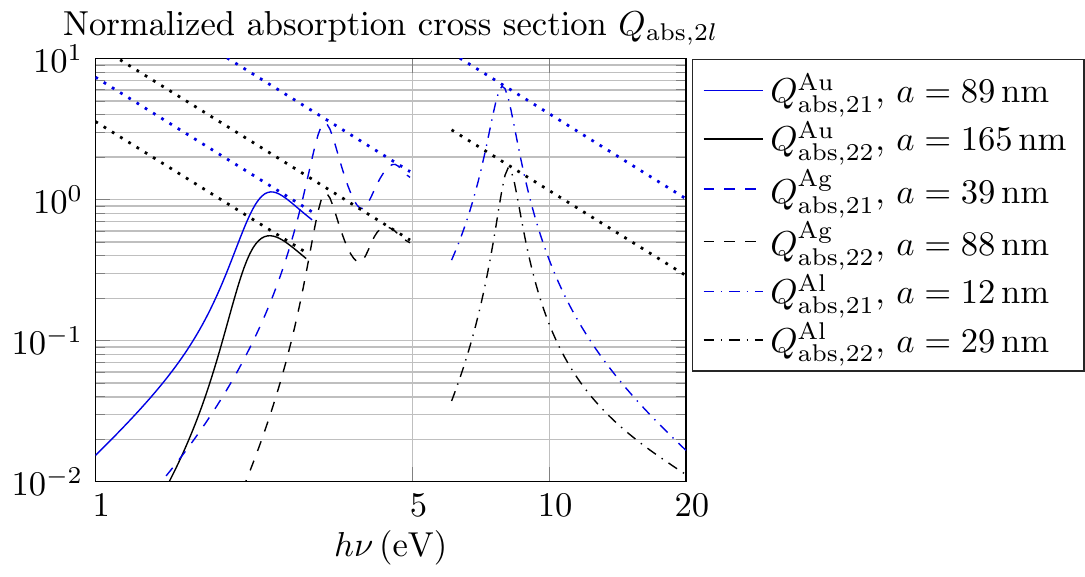}
\end{center}
\vspace{-5mm}
\caption{Normalized dipole and quadrupole absorption cross sections $Q_{\mrm{abs},2l}^\mrm{Ax}$, $l=1,2$, for a sphere made of gold (Au), silver (Ag) and aluminum (Al),
and where the radius $a$ has been tuned to match the condition \eqref{eq:opttunedcond} for optimal absorption, see also Fig.~\ref{fig:matfig45}.
The almost tangent optimal bounds are plotted as dotted lines.
Here, the background is lossless with $\epsilon_\mrm{b}=1$.}
\label{fig:matfig102}
\end{figure}

\begin{figure}[htb]
\begin{center}
\includegraphics[width=0.48\textwidth]{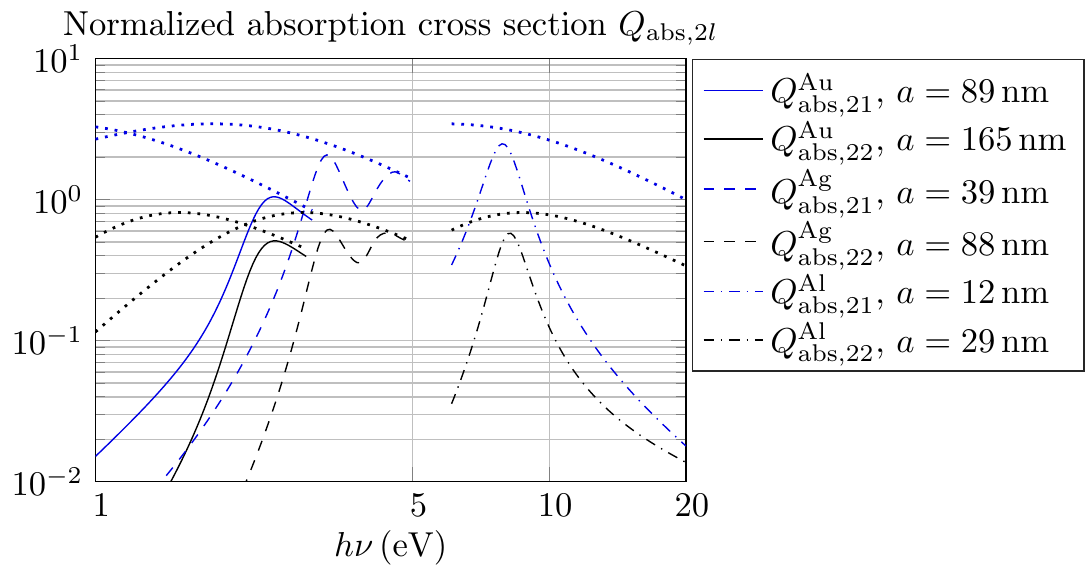}
\end{center}
\vspace{-5mm}
\caption{Same plot and parameter choices as in Fig.~\ref{fig:matfig102}, except here with a lossy background given by $\epsilon_\mrm{b}=1+\iu 0.1$.}
\label{fig:matfig103}
\end{figure}

To illustrate the theory on optimal resonances with an application in plasmonics at optical frequencies, we investigate the optimal absorption in gold (Au), silver (Ag) and aluminum (Al)
nanospheres embedded in a lossy medium as indicated above.
In Figure \ref{fig:matfig21} is plotted the permittivities of the three metals 
according to the Brendel-Bormann (BB) model fitted to experimental data as presented in 
\cite[the dielectric model in Eq. (11) with parameter values from Table 1 and Table 3]{Rakic+etal1998}.
Here, the frequency axis is given in terms of the photon energy $h\nu$ in units of electron volts\unit{(eV)} where $h$ is Planck's constant and $\nu$ the frequency.

Let $\widehat{\epsilon}_{\mrm{opt},1}(k_0a)$ and $\widehat{\epsilon}_{\mrm{opt},2}(k_0a)$ denote the approximate asymptotic expressions corresponding to 
the optimal permittivities given by \eqref{eq:poleasexprdipoleopt} and \eqref{eq:poleasexprquadrupoleopt} up to order 3 and 5, respectively.
One can now attempt to numerically solve the parametric equation
\begin{equation}\label{eq:opttunedcond}
\widehat{\epsilon}_{\mrm{opt},l}(k_0a)=\epsilon^\mrm{Ax}(\nu),
\end{equation}
for $l=1,2$, $\mrm{Ax}=\mrm{Au},\mrm{Ag},\mrm{Al}$, and where $\epsilon^\mrm{Ax}(\nu)$ denotes either of the dielectric BB-models for gold, silver and aluminum, respectively.
%An illustration of the numerical search is given in Figure \ref{fig:matfig45}.
When a solution is obtained in terms of $(k_0a,\nu)$, the optimally tuned radius is given simply as $a=k_0a \mrm{c_0}/2\pi\nu$.
It turns out that such a solution can be found for the dipole as well as for the quadrupole for all the three metals, as illustrated in Figure \ref{fig:matfig45}.
In Figure \ref{fig:matfig102} is plotted and summarized the results for the optimally tuned metal spheres for a lossless background with $\epsilon_\mrm{b}=1$,
and where $Q_{\mrm{abs},2l}^\mrm{Ax}$, $l=1,2$, is given by \eqref{eq:Qabssphexpdef} based on the corresponding BB-model for the dielectric function of gold, 
silver and aluminum, respectively. 
The optimally tuned radii appear in the legends to the right, and the corresponding (almost tangent) optimal bounds \eqref{eq:Cabsopt} are plotted as dotted lines.
As a reference, the dipole resonance peaks for gold, silver and aluminum shown in Figure \ref{fig:matfig102} appear approximately at the 
wavelengths $\lambda=510$\unit{nm},  $\lambda=390$\unit{nm} and $\lambda=158$\unit{nm}, respectively.
In Figure \ref{fig:matfig103} is illustrated the impact of a significant increase of the external losses by changing the background permittivity to 
$\epsilon_\mrm{b}=1+\iu 10^{-1}$ ($\alpha^{-1}\approx10^{-4}$ \unit{cm}), and where all the other parameters are left unchanged.
It is observed that significant losses are needed to impact on these near-optimal resonances. 
As \eg with $\epsilon_\mrm{b}=1+\iu 10^{-3}$ ($\alpha^{-1}\approx10^{-2}$ \unit{cm}), the absorption is virtually not affected at all.
These characteristics are also readily understood in view of the plots in Figure \ref{fig:matfig101}, considering that the radii of the optimal metal spheres investigated here are rather large
with an electrical size at resonance ranging from $k_0a=1/2$ to $k_0a=2$.

It should be noted that the optimization procedure described above in \eqref{eq:opttunedcond} is only sub-optimal in the sense that it is based on the asymptotic expansions
\eqref{eq:poleasexprdipoleopt} and \eqref{eq:poleasexprquadrupoleopt} rather than the exact pole positions ${\epsilon}_{\mrm{p},l}(k_0a)$ defined in Section \ref{sect:plasmonsingularity}.
Of course, this deficiency could be remedied by using a more sophisticated numerical procedure to find the maximizing permittivity function. 
However, it should also be emphasized that this example merely illustrates the fact that a gold, silver or aluminum sphere under a certain BB-model, 
can yield optimal absorption at a certain size at a certain frequency. For this particular size and frequency, there are no other materials that can give higher
absorption. But this also means \eg that if the size of the optimal multipole Ax sphere is just slightly decreased
then both $Q_{\mrm{abs},2l}^\mrm{opt}$ as well as $Q_{\mrm{abs},2l}^\mrm{Ax}$ (still close to resonance) will slightly increase, but optimality is lost for all frequencies.
This behavior is quite natural since the sphere is a very simple geometry with only one geometrical degree of freedom (its radius), whereas more complicated objects
such as the spheroid having two geometrical degrees of freedom (size and eccentricity) would be expected to give much higher flexibility in this regard.

\section{Summary and conclusions}
Fundamental upper bounds on absorbed and scattered powers are given for the plasmonic multipole resonances of a rotationally invariant sphere embedded in a lossy medium
and an asymptotic analysis is carried out to characterize the corresponding resonances of small homogeneous spheres. 
Explicit expressions are given
for the dipole and the quadrupole and the theory is illustrated in a comparison with the corresponding resonances of metal nanospheres based
on a specific Brendel-Bormann (BB) model for the permittivity of gold, silver and aluminum. 
%The theory provides explicit analytical results for a canonical problem that can be used as a benchmark and for comparisons in the development 
%of new and more general bounds and applications in this field. 

\begin{acknowledgments}
This work has been partly supported by the Swedish Foundation for Strategic Research (SSF) under the programme 
Applied Mathematics and the project Complex Analysis and Convex Optimization for EM Design.
\end{acknowledgments}

\appendix
\section{Spherical vector waves}\label{sect:spherical}
\subsection{Definition of spherical vector waves}\label{sect:sphericaldef}
In a source-free homogeneous and isotropic medium the electromagnetic field can be expanded in spherical vector waves as
\begin{equation}\label{eq:EHsphdef}
\left\{\begin{array}{l}
\bm{E}(\bm{r})=\displaystyle\sum_{l,m,\tau}a_{\tau ml}{\bf v}_{\tau ml}(k\bm{r})+f_{\tau ml}{\bf u}_{\tau ml}(k\bm{r}), \vspace{0.2cm} \\
\bm{H}(\bm{r})=\displaystyle\frac{1}{\iu\eta_0\eta}\sum_{l,m,\tau}a_{\tau ml}{\bf v}_{\bar{\tau} ml}(k\bm{r})+f_{\tau ml}{\bf u}_{\bar{\tau} ml}(k\bm{r}),  
\end{array}\right.
\end{equation}
where ${\bf v}_{\tau ml}(k\bm{r})$ and ${\bf u}_{\tau ml}(k\bm{r})$ are the regular and the outgoing spherical vector waves, respectively, 
and $a_{\tau ml}$ and $f_{\tau ml}$ the corresponding multipole coefficients,
see \eg \cite{Newton1982,Bohren+Huffman1983,Arfken+Weber2001,Jackson1999,Kristensson2016}.
Here, $l=1,2,\ldots,$ is the multipole order, $m=-l,\ldots,l$, the azimuthal index and $\tau=1,2$, where
$\tau=1$ indicates a transverse electric (\textrm{TE}) magnetic multipole and $\tau=2$ a transverse magnetic (\textrm{TM}) electric multipole,
and $\bar{\tau}$ denotes the dual index, \ie $\bar{1}=2$ and $\bar{2}=1$.

The solenoidal (source-free) regular spherical vector waves are defined here by
\begin{multline}\label{eq:v1def}
\displaystyle{\bf v}_{1 ml}(k{\bm{r}})  =   \frac{1}{\sqrt{l(l+1)}}\nabla\times({\bm{r}}\mrm{j}_l(kr)\mrm{Y}_{ml}(\hat{\bm{r}})) \\
=   \mrm{j}_l(kr){\bf A}_{1 ml}(\hat{\bm{r}}),
\end{multline}
and 
\begin{multline}\label{eq:v2def}
{\bf v}_{2 ml}(k\bm{r})   =   \displaystyle \frac{1}{k}\nabla\times{\bf v}_{1 ml}(k\bm{r}) \\
 =\displaystyle\frac{(kr\mrm{j}_l(kr))^{\prime}}{kr}{\bf A}_{2 ml}(\hat{\bm{r}})+\sqrt{l(l+1)}\frac{\mrm{j}_l(kr)}{kr}{\bf A}_{3 ml}(\hat{\bm{r}}), 
\end{multline}
where $\mrm{Y}_{ml}(\hat{\bm{r}})$ are the spherical harmonics, ${\bf A}_{\tau ml}(\hat{\bm{r}})$ the vector spherical harmonics and $\mrm{j}_l(x)$ the spherical Bessel functions of order $l$,
\cf \cite{Bostrom+Kristensson+Strom1991,Arfken+Weber2001,Jackson1999,Newton2002,Olver+etal2010,Kristensson2016}. 
Here, $(\cdot)^\prime$ denotes a differentiation with respect to the argument of the spherical Bessel function.
The outgoing (radiating) spherical vector waves ${\bf u}_{\tau ml}(k{\bm{r}})$ are obtained by replacing
the regular spherical Bessel functions $\mrm{j}_l(x)$ above with the spherical Hankel functions of the first kind, $\mrm{h}_l^{(1)}(x)$, 
see \cite{Bostrom+Kristensson+Strom1991,Jackson1999,Olver+etal2010,Kristensson2016}.
%Any one of the spherical vector waves ${\bf w}_{\tau ml}(k\bm{r})$ defined above satisfy the following curl properties
%\begin{equation}\label{eq:w12cross}
%\nabla\times {\bf w}_{\tau ml}(k\bm{r})=k{\bf w}_{\bar{\tau} ml}(k\bm{r}), \quad \tau=1,2,
%\end{equation}
%and hence the source-free Maxwell's equations (vector Helmholtz equation) in free space, \ie 
%\begin{equation}
%\nabla\times\nabla\times {\bf w}_{\tau ml}(k\bm{r})=k^2 {\bf w}_{\tau ml}(k\bm{r}),\quad \tau=1,2.
%\end{equation}

The vector spherical harmonics ${\bf A}_{\tau ml}(\hat{\bm{r}})$ are given by
\begin{equation}\label{eq:Adef}
\left\{\begin{array}{lll}
{\bf A}_{1ml}(\hat{\bm{r}})  &  =   & \displaystyle\frac{1}{\sqrt{l(l+1)}}\nabla\times\left( \bm{r}\mrm{Y}_{ml}(\hat{\bm{r}}) \right) \\
     &    =  &  \displaystyle\frac{1}{\sqrt{l(l+1)}}\nabla\mrm{Y}_{ml}(\hat{\bm{r}})\times\bm{r},   \vspace{0.2cm}\\
{\bf A}_{2ml}(\hat{\bm{r}})  &    =  &   \hat{\bm{r}}\times{\bf A}_{1ml}(\hat{\bm{r}})  \vspace{0.2cm}\\
     &  =  &   \displaystyle\frac{1}{\sqrt{l(l+1)}}r\nabla\mrm{Y}_{ml}(\hat{\bm{r}}),  \vspace{0.2cm}\\
{\bf A}_{3ml}(\hat{\bm{r}})  &  =  &  \hat{\bm{r}}\mrm{Y}_{ml}(\hat{\bm{r}}), 
\end{array}\right.
\end{equation}
where $\tau=1,2,3$, and where the spherical harmonics $\mrm{Y}_{ml}(\hat{\bm{r}})$ are given by
\begin{equation}
\mrm{Y}_{ml}(\hat{\bm{r}})=(-1)^m\sqrt{\frac{2l+1}{4\pi}}\sqrt{\frac{(l-m)!}{(l+m)!}}\mrm{P}_{l}^m(\cos\theta)\eu^{{\rm i}m\phi}, 
\end{equation}
and where $\mrm{P}_{l}^m(x)$ are the associated Legendre functions \cite{Arfken+Weber2001,Jackson1999,Olver+etal2010}.
The associated Legendre functions can be obtained from
\begin{equation}
\mrm{P}_l^m(\cos\theta)=(-1)^m(\sin\theta)^m\frac{\mrm{d}^m}{\mrm{d}(\cos\theta)^m}\mrm{P}_l(\cos\theta),   
\end{equation}
where $\mrm{P}_l(x)$ are the Legendre polynomials of order $l$ and $0\leq m\leq l$,  see \cite{Arfken+Weber2001,Jackson1999,Olver+etal2010}.
Important symmetry properties are $\mrm{P}_l^{-m}(x)=(-1)^m\frac{(l-m)!}{(l+m)!}\mrm{P}_l^m(x)$ and 
$\mrm{Y}_{-m,l}(\theta,\phi)=(-1)^m\mrm{Y}_{ml}^*(\theta,\phi)$ where $m\geq 0$. Hence, the vector spherical harmonics satisfy the symmetry 
${\bf A}_{\tau, -m,l}(\hat{\bm{r}})=(-1)^m{\bf A}^{*}_{\tau ml}(\hat{\bm{r}})$.
%where the dagger notation $(\cdot)^{\dagger}$ denotes a sign-shift in the azimuthal $m$-index.
The vector spherical harmonics are orthonormal on the unit sphere, and hence
\begin{equation}\label{eq:Aorthonormal}
\int_{\Omega_0}{\bf A}_{\tau ml}^*(\hat{\bm{r}})\cdot{\bf A}_{\tau^\prime m^\prime l^\prime }(\hat{\bm{r}})\mrm{d}\Omega
=\delta_{\tau\tau^\prime}\delta_{mm^\prime}\delta_{ll^\prime},
\end{equation}
where $\Omega_0$ denotes the unit sphere and $\mrm{d}\Omega=\sin\theta\mrm{d}\theta\mrm{d}\phi$. 

\subsection{Sum identities for the vector spherical harmonics}\label{sect:sumidspheharm}
General sum identities for the vector spherical harmonics are derived below.
We start with the addition theorem for the Legendre polynomials given by
\begin{equation}\label{eq:Pladdtheorem}
\mrm{P}_l(\hat{\bm{r}}_1\cdot\hat{\bm{r}}_2)=\frac{4\pi}{2l+1}\sum_{m=-l}^l\mrm{Y}_{ml}^*(\hat{\bm{r}}_1)\mrm{Y}_{ml}(\hat{\bm{r}}_2),
\end{equation}
see \eg \cite[Appendix C.5 on pp.~635--637]{Kristensson2016} or \cite[Eq.~(8.189) on p.~556]{Arfken+Weber2001}. 
In particular, for $\hat{\bm{r}}=\hat{\bm{r}}_1=\hat{\bm{r}}_2$ this relation reads
\begin{equation}\label{eq:Ymladdtheorem}
\sum_{m=-l}^l\mrm{Y}_{ml}^*(\hat{\bm{r}})\mrm{Y}_{ml}(\hat{\bm{r}})=\frac{2l+1}{4\pi}, \quad l=0,1,2,\ldots,
\end{equation}
since $\mrm{P}_l(1)=1$.
Notice that the sum is independent of the direction $\hat{\bm{r}}$.

Now, differentiate \eqref{eq:Pladdtheorem} to obtain to following dyadic identity
\begin{equation}\label{eq:diffPladdtheorem}
\nabla_2\nabla_1 \mrm{P}_l(\hat{\bm{r}}_1\cdot\hat{\bm{r}}_2)=\frac{4\pi}{2l+1}\sum_{m=-l}^l\nabla_2\mrm{Y}_{ml}^*(\hat{\bm{r}}_2)\nabla_1\mrm{Y}_{ml}(\hat{\bm{r}}_1),
\end{equation}
and notice that the result is not symmetric in the indices 1 and 2. The left-hand side of the identity in \eqref{eq:diffPladdtheorem} 
can be evaluated by using the differential rules of the nabla operator
\begin{multline}\label{eq:diffPladdtheoremresult}
\nabla_2\nabla_1 \mrm{P}_l(\hat{\bm{r}}_1\cdot\hat{\bm{r}}_2)
%=\nabla_2\left\{ \mrm{P}_l^\prime(\hat{\bm{r}}_1\cdot\hat{\bm{r}}_2) \nabla_1(\hat{\bm{r}}_1\cdot\hat{\bm{r}}_2)\right\} \\
=\nabla_2\left\{ \mrm{P}_l^\prime(\hat{\bm{r}}_1\cdot\hat{\bm{r}}_2) \left(\frac{\hat{\bm{r}}_2}{r_1}-\hat{\bm{r}}_1\frac{\hat{\bm{r}}_1\cdot\hat{\bm{r}}_2}{r_1} \right) \right\} \\
=\mrm{P}_l^{\prime\prime}(\hat{\bm{r}}_1\cdot\hat{\bm{r}}_2)\left(\frac{\hat{\bm{r}}_1}{r_2}-\hat{\bm{r}}_2\frac{\hat{\bm{r}}_1\cdot\hat{\bm{r}}_2}{r_2} \right)
\left(\frac{\hat{\bm{r}}_2}{r_1}-\hat{\bm{r}}_1\frac{\hat{\bm{r}}_1\cdot\hat{\bm{r}}_2}{r_1} \right) \\
+\mrm{P}_l^\prime(\hat{\bm{r}}_1\cdot\hat{\bm{r}}_2)\left(\frac{{\bf I}_3-\hat{\bm{r}}_2\hat{\bm{r}}_2}{r_1r_2} 
-\frac{\hat{\bm{r}}_1-\hat{\bm{r}}_2(\hat{\bm{r}}_1\cdot\hat{\bm{r}}_2)}{r_1r_2}\hat{\bm{r}}_1
\right).
\end{multline}
In the derivation of \eqref{eq:diffPladdtheoremresult} we have used 
\begin{equation}
\nabla_1(\hat{\bm{r}}_1\cdot\hat{\bm{r}}_2)
%=\nabla_1\frac{\bm{r}_1\cdot\hat{\bm{r}}_2}{r_1}
%=\frac{\hat{\bm{r}}_2}{r_1}-\hat{\bm{r}}_1\frac{\bm{r}_1\cdot\hat{\bm{r}}_2}{r_1^2}
=\frac{\hat{\bm{r}}_2}{r_1}-\hat{\bm{r}}_1\frac{\hat{\bm{r}}_1\cdot\hat{\bm{r}}_2}{r_1} 
\end{equation}
and similarly for $\nabla_2(\hat{\bm{r}}_1\cdot\hat{\bm{r}}_2)$, as well as
\begin{multline}
\nabla_2\left(\frac{\hat{\bm{r}}_2}{r_1}-\hat{\bm{r}}_1\frac{\hat{\bm{r}}_1\cdot\hat{\bm{r}}_2}{r_1} \right)
%=\frac{\nabla_2\hat{\bm{r}}_2}{r_1}-\frac{\hat{\bm{r}}_1-\hat{\bm{r}}_2(\hat{\bm{r}}_1\cdot\hat{\bm{r}}_2)}{r_1r_2}\hat{\bm{r}}_1 \\
=\frac{{\bf I}_3-\hat{\bm{r}}_2\hat{\bm{r}}_2}{r_1r_2} -\frac{\hat{\bm{r}}_1-\hat{\bm{r}}_2(\hat{\bm{r}}_1\cdot\hat{\bm{r}}_2)}{r_1r_2}\hat{\bm{r}}_1 
\end{multline}
and where
\begin{equation}
\nabla\hat{\bm{r}}
%=\nabla\frac{\bm{r}}{r}=\frac{{\bf I}_3}{r}-\frac{\hat{\bm{r}}\bm{r}}{r^2}
=\frac{{\bf I}_3-\hat{\bm{r}}\hat{\bm{r}}}{r},
\end{equation}
and ${\bf I}_3$ is the identity dyadic.
The result \eqref{eq:diffPladdtheoremresult} can now be simplified and combined with the right-hand side of \eqref{eq:diffPladdtheorem} to yield
\begin{multline}\label{eq:sumrelationr1r2Yml}
r_1r_2\nabla_2\nabla_1 \mrm{P}_l(\hat{\bm{r}}_1\cdot\hat{\bm{r}}_2) 
%=\mrm{P}_l^{\prime\prime}(\hat{\bm{r}}_1\cdot\hat{\bm{r}}_2)\left(\hat{\bm{r}}_1-\hat{\bm{r}}_2(\hat{\bm{r}}_1\cdot\hat{\bm{r}}_2) \right)
%\left(\hat{\bm{r}}_2-\hat{\bm{r}}_1(\hat{\bm{r}}_1\cdot\hat{\bm{r}}_2) \right) \\
%+\mrm{P}_l^\prime(\hat{\bm{r}}_1\cdot\hat{\bm{r}}_2)\left({\bf I}_3-\hat{\bm{r}}_2\hat{\bm{r}}_2 
%-(\hat{\bm{r}}_1-\hat{\bm{r}}_2(\hat{\bm{r}}_1\cdot\hat{\bm{r}}_2))\hat{\bm{r}}_1\right) \\
=\mrm{P}_l^{\prime\prime}(\hat{\bm{r}}_1\cdot\hat{\bm{r}}_2)
\left( \hat{\bm{r}}_1\hat{\bm{r}}_2-\hat{\bm{r}}_1\hat{\bm{r}}_1(\hat{\bm{r}}_1\cdot\hat{\bm{r}}_2) \right. \\
\left. -\hat{\bm{r}}_2\hat{\bm{r}}_2(\hat{\bm{r}}_1\cdot\hat{\bm{r}}_2)+\hat{\bm{r}}_2\hat{\bm{r}}_1(\hat{\bm{r}}_1\cdot\hat{\bm{r}}_2)^2\right) \\
+\mrm{P}_l^\prime(\hat{\bm{r}}_1\cdot\hat{\bm{r}}_2)\left({\bf I}_3-\hat{\bm{r}}_2\hat{\bm{r}}_2 
-\hat{\bm{r}}_1\hat{\bm{r}}_1+\hat{\bm{r}}_2\hat{\bm{r}}_1(\hat{\bm{r}}_1\cdot\hat{\bm{r}}_2)\right) \\
=\frac{4\pi}{2l+1}\sum_{m=-l}^l r_2\nabla_2\mrm{Y}_{ml}^*(\hat{\bm{r}}_2)r_1\nabla_1\mrm{Y}_{ml}(\hat{\bm{r}}_1).
\end{multline}
In particular, for $\hat{\bm{r}}=\hat{\bm{r}}_1=\hat{\bm{r}}_2$ the relation \eqref{eq:sumrelationr1r2Yml} reads
\begin{equation}\label{eq:sumrelationrYml}
\sum_{m=-l}^l \left(r\nabla\mrm{Y}_{ml}^*(\hat{\bm{r}})\right)\left(r\nabla\mrm{Y}_{ml}(\hat{\bm{r}})\right)=\frac{2l+1}{4\pi}\frac{l(l+1)}{2}\left( {\bf I}_3-\hat{\bm{r}}\hat{\bm{r}}\right)
\end{equation}
since $\mrm{P}_l^\prime(1)=l(l+1)/2$.

By employing the definitions made in \eqref{eq:Adef} we see that \eqref{eq:sumrelationrYml} can be written
\begin{equation}\label{eq:sumA2}
\sum_{m=-l}^l{\bf A}_{2ml}^*(\hat{\bm{r}}){\bf A}_{2ml}(\hat{\bm{r}})=\frac{2l+1}{8\pi}\left( {\bf I}_3-\hat{\bm{r}}\hat{\bm{r}}\right).
\end{equation}
Similarly, 
\begin{multline}\label{eq:sumA1}
\sum_{m=-l}^l{\bf A}_{1ml}^*(\hat{\bm{r}}){\bf A}_{1ml}(\hat{\bm{r}})=
-\hat{\bm{r}}\times\sum_{m=-l}^l{\bf A}_{2ml}^*(\hat{\bm{r}}){\bf A}_{2ml}(\hat{\bm{r}})\times\hat{\bm{r}} \\
=-\frac{2l+1}{8\pi}\hat{\bm{r}}\times\left( {\bf I}_3-\hat{\bm{r}}\hat{\bm{r}}\right)\times\hat{\bm{r}}
=\frac{2l+1}{8\pi}\left( {\bf I}_3-\hat{\bm{r}}\hat{\bm{r}}\right),
\end{multline}
and from \eqref{eq:Ymladdtheorem} follows that
\begin{equation}\label{eq:sumA3}
\sum_{m=-l}^l{\bf A}_{3ml}^*(\hat{\bm{r}}){\bf A}_{3ml}(\hat{\bm{r}})=\frac{2l+1}{4\pi}\hat{\bm{r}}\hat{\bm{r}}.
\end{equation}
Finally, by adding over all indices $\tau=1,2,3$ we obtain the result
\begin{equation}\label{eq:sumA123}
\sum_{m=-l}^l \sum_{\tau=1}^3{\bf A}_{\tau ml}^*(\hat{\bm{r}}){\bf A}_{\tau ml}(\hat{\bm{r}})=\frac{2l+1}{4\pi}{\bf I}_3,
\end{equation}
which is independent of the direction $\hat{\bm{r}}$.

%\subsection{First Lommel integral for spherical Bessel functions with complex-valued argument}\label{sect:Lommel}
\subsection{Lommel integrals for spherical Bessel functions}\label{sect:Lommel}
Let ${\rm s}_l(kr)$ denote an arbitrary linear combination of spherical Bessel and Hankel functions.
Based on the two Lommel integrals for cylinder functions, \cf \cite[Eqs.~(10.22.4) and (10.22.5) on p.~241]{Olver+etal2010} and \cite[Eqs.~(8) and (10) on p.~134]{Watson1995},
the following indefinite Lommel integrals can be derived for spherical Bessel functions
\begin{equation}\label{eq:firstLommel}
\int\left|{\rm s}_l(kr)\right|^2r^2\mrm{d} r
=r^2\frac{ \Im\!\left\{k{\rm s}_{l+1}(kr) {\rm s}_{l}^*(kr) \right\}}{\Im\!\left\{k^2\right\}},
\end{equation}
where $k$ is complex-valued ($k\neq k^*$), \cf \cite[Eq.~(A.15) on p.~11]{Nordebo+etal2017a}, and
\begin{multline}\label{eq:secondLommel}
\int\left|{\rm s}_l(kr)\right|^2r^2\mrm{d} r \\
=\frac{1}{2}r^3\left(\left| {\rm s}_l(kr) \right|^2-\Re\{{\rm s}_{l-1}(kr){\rm s}_{l+1}^*(kr)\} \right),
\end{multline}
where $k$ is real-valued ($k=k^*$).
Furthermore, by using the recursive relationships
\begin{equation}\label{eq:recursive}
\left\{\begin{array}{l}
\displaystyle \frac{\mrm{s}_l(kr)}{kr}=\frac{1}{2l+1}\left(\mrm{s}_{l-1}(kr)+\mrm{s}_{l+1}(kr) \right), \vspace{0.2cm} \\
\displaystyle \mrm{s}_l^\prime(kr)=\frac{1}{2l+1}\left(l\mrm{s}_{l-1}(kr)-(l+1)\mrm{s}_{l+1}(kr) \right), 
\end{array}\right.
\end{equation}
where $l=1,2,\ldots$, \cf \cite{Olver+etal2010}, it can be shown that
\begin{multline}\label{eq:recursiveLommel}
\int\left(\left|\frac{\mrm{s}_{l}(kr)}{kr}+\mrm{s}_{l}^\prime(kr)\right|^2 
 +l(l+1)  \left|\frac{\mrm{s}_{l}(kr)}{kr}\right|^2\right)r^2\mrm{d} r  \\
 =\frac{1}{2l+1}\int \left( (l+1)\left|{\rm s}_{l-1}(kr)\right|^2+l\left|{\rm s}_{l+1}(kr)\right|^2\right)r^2\mrm{d} r, 
\end{multline}
see also \eg \cite[Eq.~(17) on p.~411]{Marengo+Devaney1999} and \cite[Eqs.~(36) and (47) on pp.~2359--2360]{Nordebo+Gustafsson2006}.

\subsection{Orthogonality over a spherical volume}\label{sect:sphericalorthvol}
Due to the orthonormality of the vector spherical harmonics \eqref{eq:Aorthonormal},
the regular spherical vector waves are orthogonal over the unit sphere with
\begin{equation}\label{eq:vorthogonal1}
\displaystyle\int_{\Omega_0}{\bf v}_{\tau ml}^*(k{\bm{r}})\cdot{\bf v}_{\tau^\prime m^\prime l^\prime}(k{\bm{r}})\mrm{d} \Omega 
=\displaystyle\delta_{\tau\tau^\prime}\delta_{mm^\prime}\delta_{ll^\prime}S_{\tau l}(k,r), 
\end{equation}
where
\begin{multline}\label{eq:Stauldef}
S_{\tau l}(k,r)=\displaystyle\int_{\Omega_0}|{\bf v}_{\tau ml}(k\bm{r})|^2\mrm{d}\Omega  \\
 =\left\{\begin{array}{ll}
\displaystyle  \left|\mrm{j}_{l}(kr)\right|^2 &   \tau=1,  \vspace{0.2cm} \\
\displaystyle  \left|\frac{\mrm{j}_{l}(kr)}{kr}+\mrm{j}_{l}^\prime(kr)\right|^2+l(l+1) \left|\frac{\mrm{j}_{l}(kr)}{kr}\right|^2  &   \tau=2. 
\end{array}\right.
\end{multline}
As a consequence, the regular spherical vector waves are also orthogonal over a spherical volume $\mrm{V}_a$ with radius $a$ yielding
\begin{equation}\label{eq:vorthogonal2}
\displaystyle\int_{\mrm{V}_a}{\bf v}_{\tau ml}^*(k{\bm{r}})\cdot{\bf v}_{\tau^\prime m^\prime l^\prime}(k{\bm{r}})\mrm{d} v 
=\displaystyle\delta_{\tau\tau^\prime}\delta_{mm^\prime}\delta_{ll^\prime}W_{\tau l}(k,a), 
\end{equation}
where
\begin{equation}\label{eq:Wtauldef}
W_{\tau l}(k,a)=\int_{\mrm{V}_a}\left|{\bf v}_{\tau ml}(k\bm{r})\right|^2\mrm{d} v 
=\int_{0}^{a}S_{\tau l}(k,r)r^2\mrm{d} r, 
\end{equation}
where $\mrm{d} v=r^2\mrm{d}\Omega\mrm{d} r$ and $\tau=1,2$.

For complex-valued arguments $k$, $W_{1l}(k,a)$ is obtained from \eqref{eq:firstLommel} as
\begin{equation}\label{eq:W1ldef}
W_{1l}(k,a)=\int_{0}^{a}\left|\mrm{j}_{l}(kr)\right|^2 r^2\mrm{d} r 
=\frac{a^2 \Im\!\left\{k\mrm{j}_{l+1}(ka) \mrm{j}_{l}^*(ka) \right\}}{\Im\!\left\{k^2\right\}}, 
\end{equation}
and from \eqref{eq:recursiveLommel} follows that
\begin{multline}\label{eq:W2ldef}
W_{2 l}(k,a) \\
=\int_{0}^{a}\left(\left|\frac{\mrm{j}_{l}(kr)}{kr}+\mrm{j}_{l}^\prime(kr)\right|^2 
 +l(l+1)  \left|\frac{\mrm{j}_{l}(kr)}{kr}\right|^2\right)r^2\mrm{d} r  \\
 =\frac{1}{2l+1}\left((l+1)W_{1,l-1}(k,a)+lW_{1,l+1}(k,a) \right).  
\end{multline}
Similar expressions are obtained for real-valued $k$ by using \eqref{eq:secondLommel}. 
%The relations above can also be used to evaluate volume integrals over of the outgoing spherical vector waves based on 
% spherical Hankel functions $\mrm{h}_{l}^{(1)}(ka)$ over spherical shells $0<a\leq r\leq R<\infty$.

\subsection{Orthogonality over a spherical surface}\label{sect:sphericalorthsurf}
Based on the properties of the spherical vector waves described in Section \ref{sect:sphericaldef}, the
following orthogonality relationships regarding their cross products on a spherical surface can be derived%{eq:v1def}{eq:Adef}{eq:Aorthonormal}
\begin{multline}\label{eq:orthcrosssph1}
\int_{\mrm{S}_a}{\bf w}_{\tau ml}(k\bm{r})\times {\bf z}_{\bar{\tau} m^\prime l^\prime}^*(k\bm{r})\cdot\mrm{d}\bm{S} \\
=a^2\delta_{mm^\prime}\delta_{ll^\prime}
\left\{\begin{array}{ll}
 \displaystyle  w_l(ka)\left(\frac{(ka z_l(ka))^\prime}{ka} \right)^*  & \tau=1, \vspace{0.2cm} \\
\displaystyle  -\left(\frac{(ka w_l(ka))^\prime}{ka} \right)z_l^*(ka) & \tau=2,
\end{array}\right.
\end{multline}
and 
\begin{equation}\label{eq:orthcrosssph2}
\int_{\mrm{S}_a}{\bf w}_{\tau ml}(k\bm{r})\times {\bf z}_{\tau m^\prime l^\prime}^*(k\bm{r})\cdot\mrm{d}\bm{S}=0,
\end{equation}
for $\tau=1,2$.
Here, $\mrm{S}_a$ is the spherical surface of radius $a$, and $w_l(ka)$ and $z_l(ka)$ are either of $\mrm{j}_{l}(ka)$ or $\mrm{h}_{l}^{(1)}(ka)$ 
and ${\bf w}_{\tau ml}(k\bm{r})$ and ${\bf z}_{\tau ml}(k\bm{r})$ the corresponding spherical vector waves, respectively.

%%\bibliography{apssamp}% Produces the bibliography via BibTeX.
%\bibliographystyle{unsrt}
%\bibliography{total}

\begin{thebibliography}{10}

\bibitem{Maier2007}
S.~A. Maier.
\newblock {\em Plasmonics: Fundamentals and Applications}.
\newblock Sprin\-ger-Verlag, Berlin, 2007.

\bibitem{Valagiannopoulos+etal2015}
C.~A. Valagiannopoulos, J.~Vehmas, C.~R. Simovski, S.~A. Tretyakov, and S.~I.
  Maslovski.
\newblock Electromagnetic energy sink.
\newblock {\em Phys. Rev. B}, 92:245402, 2015.

\bibitem{Maslovski+etal2016}
S.~I. Maslovski, C.~R. Simovski, and S.~A. Tretyakov.
\newblock Overcoming black body radiation limit in free space: metamaterial
  superemitter.
\newblock {\em New J. Phys.}, 18:013034, 2016.

\bibitem{Valagiannopoulos+Tretyakov2016}
C.~A. Valagiannopoulos and S.~A. Tretyakov.
\newblock Theoretical concepts of unlimited-power reflectors, absorbers, and
  emitters with conjugately matched layers.
\newblock {\em Phys. Rev. B}, 94:125117, 2016.

\bibitem{Bohren+Huffman1983}
C.~F. Bohren and D.~R. Huffman.
\newblock {\em Absorption and Scattering of Light by Small Particles}.
\newblock John Wiley \& Sons, New York, 1983.

\bibitem{Progelhof+etal1971}
R.C. Progelhof, J.~Franey, and T.~W. Haas.
\newblock Absorption coefficient of unpigmented poly(methyl methacrylate),
  polystyrene, polycarbonate and poly(4-methylpentene-1) sheets.
\newblock {\em Journal of applied polymer science}, 15:1803--1807, 1971.

\bibitem{AlTaay+etal2015}
W.A. Al-Taay, S.~F. Oboudi, E.~Yousif, M.~A. Nabi, R.~M. Yusop, and D.~Derawi.
\newblock Fabrication and characterization of nickel chloride doped {PMMA}
  films.
\newblock {\em Advances in Materials Science and Engineering}, 2015:1--5, 2015.

\bibitem{Duck1990}
F.~A. Duck.
\newblock {\em Physical Properties of Tissue: A Comprehensive Reference Book}.
\newblock Academic Press, 1990.

\bibitem{Huang+etal2008}
X.~Huang, P.~K. Jain, I.~H. El-Sayed, and M.~A. El-Sayed.
\newblock Plasmonic photothermal therapy ({PPTT}) using gold nanoparticles.
\newblock {\em Lasers Med Sci}, 23:217--228, 2008.

\bibitem{Mundy+etal1974}
W.~C. Mundy, J.~A. Roux, and A.~M. Smith.
\newblock Mie scattering by spheres in an absorbing medium.
\newblock {\em J. Opt. Soc. Am.}, 64(12):1593--1597, 1974.

\bibitem{Chylek1977}
P.~Chylek.
\newblock Light scattering by small particles in an absorbing medium.
\newblock {\em J. Opt. Soc. Am.}, 67(4):561--563, 1977.

\bibitem{Bohren+Gilra1979}
C.~F. Bohren and D.~P. Gilra.
\newblock Extinction by a spherical particle in an absorbing medium.
\newblock {\em J. Colloid Interface Sci.}, 72(2):215--221, 1979.

\bibitem{Lebedev+etal1999}
A.N. Lebedev, M.~Gartz, U.~Kreibig, and O.~Stenzel.
\newblock Optical extinction by spherical particles in an absorbing medium:
  Application to composite absorbing films.
\newblock {\em Eur. Phys. J. D.}, 6(3):365--373, 1999.

\bibitem{Sudiarta+Chylek2001}
I.~W. Sudiarta and P.~Chylek.
\newblock Mie-scattering formalism for spherical particles embedded in an
  absorbing medium.
\newblock {\em J. Opt. Soc. Am. A}, 18(6):1275--1278, 2001.

\bibitem{Durant+etal2007a}
S.~Durant, O.~Calvo-Perez, N.~Vukadinovic, and J-J. Greffet.
\newblock Light scattering by a random distribution of particles embedded in
  absorbing media: diagrammatic expansion of the extinction coefficient.
\newblock {\em J. Opt. Soc. Am. A}, 24(9):2943--2952, 2007.

\bibitem{Jackson1999}
J.~D. Jackson.
\newblock {\em Classical Electrodynamics}.
\newblock John Wiley \& Sons, New York, third edition, 1999.

\bibitem{Kristensson2016}
G.~Kristensson.
\newblock {\em Scattering of Electromagnetic Waves by Obstacles}.
\newblock SciTech Publishing, Edison, NJ, 2016.

\bibitem{Osipov+Tretyakov2017}
Andrey Osipov and Sergei Tretyakov.
\newblock {\em Modern Electromagnetic Scattering Theory with Applications}.
\newblock John Wiley \& Sons Ltd, Chichester, UK, 2017.

\bibitem{Nordebo+etal2017a}
S.~Nordebo, M.~Dalarsson, Y.~Ivanenko, D.~Sj\"{o}berg, and R.~Bayford.
\newblock On the physical limitations for radio frequency absorption in gold
  nanoparticle suspensions.
\newblock {\em J. Phys. D: Appl. Phys.}, 50(15):155401, 2017.

\bibitem{Tretyakov2014}
S.~Tretyakov.
\newblock Maximizing absorption and scattering by dipole particles.
\newblock {\em Plasmonics}, 9(4):935--944, 2014.

\bibitem{Olver+etal2010}
F.~W.~J. Olver, D.~W. Lozier, R.~F. Boisvert, and C.~W. Clark.
\newblock {\em {NIST} {H}andbook of mathematical functions}.
\newblock Cambridge University Press, New York, 2010.

\bibitem{Olver1997}
F.~W.~J. Olver.
\newblock {\em Asymptotics and special functions}.
\newblock A K Peters, Ltd, Natick, Massachusetts, 1997.

\bibitem{Hormander1983}
L.~H\"ormander.
\newblock {\em The Analysis of Linear Partial Differential Operators I}.
\newblock Grundlehren der mathematischen {W}issenschaften 256.
  Sprin\-ger-Verlag, Berlin Heidelberg, 1983.

\bibitem{Tzarouchis+etal2016}
D.~C. Tzarouchis, P.~Yl\"{a}-Oijala, and A.~Sihvola.
\newblock Unveiling the scattering behavior of small spheres.
\newblock {\em Phys. Rev. B}, 94(14):140301, Oct 2016.

\bibitem{Rakic+etal1998}
A.~D. Rakic, A.~B. Djurisic, J.~M. Elazar, and M.~L. Majewski.
\newblock Optical properties of metallic films for vertical-cavity
  optoelectronic devices.
\newblock {\em Applied Optics}, 37:5271--5283, 1998.

\bibitem{Newton1982}
Roger~G. Newton.
\newblock {\em Scattering Theory of Waves and Particles}.
\newblock Sprin\-ger-Verlag, New York, 1982.

\bibitem{Arfken+Weber2001}
George~B. Arfken and Hans~J. Weber.
\newblock {\em Mathematical Methods for Physicists}.
\newblock Academic Press, New York, fifth edition, 2001.

\bibitem{Bostrom+Kristensson+Strom1991}
A.~Bostr{\"o}m, G.~Kristensson, and S.~Str{\"o}m.
\newblock Transformation properties of plane, spherical and cylindrical scalar
  and vector wave functions.
\newblock In V.~V. Varadan, A.~Lakhtakia, and V.~K. Varadan, editors, {\em
  Field Representations and Introduction to Scattering}, Acoustic,
  Electromagnetic and Elastic Wave Scattering, chapter~4, pages 165--210.
  Elsevier Science Publishers, Amsterdam, 1991.

\bibitem{Newton2002}
Roger~G. Newton.
\newblock {\em Scattering Theory of Waves and Particles}.
\newblock Dover Publications, New York, second edition, 2002.

\bibitem{Watson1995}
G.~N. Watson.
\newblock {\em A Treatise on the Theory of Bessel Functions}.
\newblock Cambridge University Press, Cambridge, U.K., second edition, 1995.

\bibitem{Marengo+Devaney1999}
E.~A. Marengo and A.~J. Devaney.
\newblock The inverse source problem of electromagnetics: Linear inversion
  formulation and minimum energy solution.
\newblock {\em IEEE Trans. Antennas Propagat.}, 47(2):410--412, February 1999.

\bibitem{Nordebo+Gustafsson2006}
Sven Nordebo and Mats Gustafsson.
\newblock Statistical signal analysis for the inverse source problem of
  electromagnetics.
\newblock {\em IEEE Trans. Signal Process.}, 54(6):2357--2361, June 2006.

\end{thebibliography}

\end{document}